\documentclass[12pt,preprint]{aastex}

\slugcomment{Submitted to {\it The Astrophysical Journal}}

\shorttitle{Chemical Abundances in Red Gaints}
\shortauthors{Smith et al.}

\begin{document}


\title{Chemical Abundances in Field Red Giants from High-Resolution
H-Band Spectra using the APOGEE Spectral Linelist}


\author{Verne V. Smith and Katia Cunha}
\affil{National Optical Astronomy Observatories, Tucson,  AZ  85719 USA and}
\affil{Observatorio Nacional, Sao Cristovao, Rio de Janeiro, Brazil}
\email{vsmith@noao.edu}

\author{Matthew D. Shetrone}
\affil{Department of Astronomy and McDonald Observatory, University of Texas, Austin,  TX 78712  USA}

\author{Szabolcs Meszaros and Carlos Allende Prieto}
\affil{Instituto d'Astrofisica de Canarias, 38205, La Laguna, Tenerife,  Spain}

\author{Dmitry Bizyaev}
\affil{Apache Point Observatory, Sunspot, NM 88349 USA and}
\affil{Strenberg Astronomical Institute, Moscow  119992, Russia}

\author{Ana Garcia Perez and Steven R. Majewski}
\affil{Department of Astronomy, University of Virginia, Charlottesville,  VA  22904 USA}


\author{Ricardo Schiavon}
\affil{Astrophysics Research Institute, Liverpool John Moores University, Liverpool L3 5UX UK}

\author{Jon Holtzman}
\affil{Department of Astronomy, New Mexico State University, Las Cruces, NM  88003 USA}

\author{Jennifer A. Johnson}
\affil{Department of Astronomy, Ohio State University, Columbus, OH  43210}




\begin{abstract}
High-resolution H-band spectra of five bright field K, M, and MS giants, 
obtained from the archives of the Kitt Peak National Observatory (KPNO)
Fourier Transform Spectrometer (FTS), are analyzed to determine chemical 
abundances of 16 elements.  The abundances were
derived via spectrum synthesis using the detailed linelist prepared for the SDSS III
Apache Point Galactic Evolution Experiment (APOGEE), which is a high-resolution
near-infrared spectroscopic survey to derive detailed chemical abundance distributions
 and precise radial velocities for 100,000 red giants sampling all Galactic stellar
populations.  The red giant sample studied here was chosen to probe which chemical
elements can be derived reliably from the H-band APOGEE spectral region.  These
red giants consist
of two K-giants ($\alpha$ Boo and $\mu$ Leo), two M-giants ($\beta$ And and
$\delta$ Oph), and one thermally-pulsing asymptotic giant branch (TP-AGB) star of
spectral type MS (HD 199799).  Measured chemical abundances include the
cosmochemically important isotopes $^{12}$C, $^{13}$C, $^{14}$N, and $^{16}$O,
along with Mg, Al, Si, K, Ca, Ti, V, Cr, Mn, Fe, Co, Ni, and Cu.  The K and M giants exhibit
the abundance signature of the first dredge-up of CN-cycle material, while the TP-AGB
star shows clear evidence of the addition of $^{12}$C synthesized during $^{4}$He-burning
thermal pulses and subsequent third dredge-up.  A comparison of the abundances derived
here with published values for these stars reveals consistent results to $\sim$0.1 dex.
The APOGEE spectral region and linelist is, thus,
well-suited for probing both Galactic chemical evolution, as well as internal
nucleosynthesis and mixing in populations of red giants via high-resolution spectroscopy.
\end{abstract}



\section{Introduction}

The Apache Point Observatory Galactic Evolution Experiment (APOGEE) 
is one of 4 experiments that are part of the Sloan Digital Sky Survey III (SDSS-III;
Eisenstein et al. 2011).
APOGEE is obtaining high-resolution (R$\sim$22,300), high signal-to-noise ratio 
(S/N$\ge$100 per pixel), H-band ($\lambda$1.51--1.69 $\mu$m) spectra 
of evolved, late-type stars, with the goal being to measure chemical abundances of
15 elements per star.  Coupled with radial velocities that are accurate to 
$\sim$100 m-s$^{-1}$, APOGEE is creating the first high-precision spectroscopic 
and radial-velocity survey of all Galactic stellar populations (bulge, bar, disks, halo) 
using a uniform set of stellar tracers and spectral diagnostics, with a plan to observe
and derive these parameters for 100,000 red giants by the end of 2014.

The physical and chemical parameters of the APOGEE stars are derived from a
suite of software packages that, together, are called the APOGEE Stellar Parameters
and Chemical Abundances Pipeline (ASPCAP); the details of this software and its
analysis techniques will be discussed by Allende-Prieto et al. (2012, in preparation).
The first-generation output of ASPCAP, which will be based on 1-D stellar atmosphere
analyses in LTE, will consist of the stellar parameters of effective temperature,
T$_{\rm eff}$, surface gravity (noted as log g), and the 1-D microturbulence
parameter ($\xi$), along with the abundances of up to 16 elements (Fe, $^{12}$C, $^{14}$N,
$^{16}$O, Mg, Al, Si, K, Ca, Ti, V, Cr, Mn, Co, Ni, or Cu) and, for many stars, 
the $^{13}$C isotopic abundance.  

One of the central components of the ASPCAP machinery is a spectral linelist 
constructed to produce the synthetic spectra that are used to compare to the
observed ones; the APOGEE linelist will be presented and discussed in detail by 
Shetrone et al. (2012, in preparation).  

 The goal of this study is to analyze a small number nearby field red giants with well-defined stellar
parameters and use the APOGEE linelist to derive detailed chemical abundances via standard ``manual''
abundance analysis techniques (Ramirez and Allende Prieto 2011 is a recent example using such techniques).
The derived stellar parameters and chemical abundances can be compared to the same quantities
derived via ASPCAP using the same spectra, linelist, and model atmosphere grid.
This comparison will be an important test of ASPCAP.  This work will also present new
abundances in these stars based on an extensive and up-to-date spectral linelist in the near-IR H-band.

Abundances will be presented in this paper using two
different types of nomenclature, with one abundance scale defined as 
A(X)=Log(N(X)/N(H)) + 12.0.  The other scale compares the element ratio of N(X)/N(Q) in a
target star to the Sun via [X/Q]=(A(X)$_{\rm Target Star}$ -- A(X)$_{\odot}$) -- (A(Q)$_{\rm Target Star}$ -- A(Q)$_{\odot}$). 

\section{Observational Data: The Fourier Transform Spectra of Bright Field Red Giants}

The APOGEE high-resolution spectra cover the wavelength range from $\sim$$\lambda$15100 to
16900\AA\ and to analyze bright red giants over this spectral interval,
data were obtained from the archives of the Fourier transform spectrometer (FTS--Hall
et al. 1979)  that operated at the coude focus of the Kitt Peak National Observatory 
Mayall 4-m reflector.  A more complete description of observing techniques and the original
data is discussed in Hinkle, Wallace, \& Livingston (1995), in particular for the spectral atlas
of $\alpha$ Boo. 

The originally reduced spectra of selected stars were recorded as flux versus wavenumber (cm$^{-1}$) 
over different dates and runs and covered
a broad region in the H-band.  The spectra analyzed here were resampled in wavelength space 
and restricted to a
wavelength range of $\lambda$15000 to 17000\AA.  The resolution of the spectra for different stars 
varies from R=45,000 to 100,000, but were all higher than what is being
observed for APOGEE.  The higher resolution allows for a detailed investigation of the limits of what can be derived
high-quality (i.e., high resolution and high signal-to-noise) red-gaint spectra when using the APOGEE linelist. 

\section{The ``Standard Star'' Parameters}

\subsection{The Choices}

There are five stars analyzed here: the well-studied $\alpha$ Boo, a mildly metal-poor
K1.5 giant, $\mu$ Leo, the prototype metal-rich star (Spinrad \& Taylor 1969), which is a K1 
giant, two near-solar-metallicity M giants, $\beta$ And (M0III) and $\delta$ Oph (M0III), and
HD199799, an AGB star of spectral type MS with a mild enhancement of $^{12}$C and a
slightly elevated ratio of C/O (relative to solar) due to third dredge-up (Smith \& Lambert 1990).
Figure 1 illustrates a small portion of the APOGEE wavelength coverage for four of the red
giants here.  This region from 15560 -- 15590\AA\ includes representative species and
how they vary with spectral type (or effective temperature) and metallicity.  Lines from the
main CNO-containing molecules, CO, OH, and CN fall within this region, as noted in the
figure.  Of particular interest is the metal-rich K-giant $\mu$ Leo, which exhibits quite
strong CN lines; much of the variation in the spectrum of $\mu$ Leo compared to the other
giants result from the high line-density of CN.  This figure also points to the desirability of
analyzing large numbers of red giants via spectrum synthesis techniques using a detailed
linelist.

With a significant fraction of its red-giant targets in the thin and thick disks, plus the
inner bulge and bar, the five types of red giants studied here span a range of temperature and
metallicity that provides a good test for ASPCAP.  The APOGEE survey targets red giants with
effective temperatures from T$_{\rm eff}$=3400K--5000K, having surface gravities from log g=3.0 to -0.5,
and metallicities from [Fe/H]= -5.0 to +1.0.
These nearby giants have  well-measured photometry and parallaxes,
so temperatures, luminosities, and masses are all rather well-constrained.   In
addition, four of the five stars have previous abundance analyses in the literature and thus provide
comparison tests for the abundances derived from the APOGEE linelist.

\subsection{Basic Data}

Deriving stellar chemical abundances from high-resolution spectra requires as basic
input the effective temperature (T$_{\rm eff}$), surface gravity (characterized as log g with g as cm-s$^{-2}$),
and overall metallicity as the fundamental parameters of the model atmosphere.  In addition,
high-resolution abundance analyses of cool stars that use static 1-D stellar atmospheres also 
require the derivation of a non-thermal Doppler-like broadening term called microturbulence ($\xi$),
which is determined as part of the spectral analysis.  The
first step is the determination of T$_{\rm eff}$ and, for these relatively nearby red giants,
is set by the near-infrared magnitudes of J and K.  This particular choice is used
because the APOGEE targets are selected from the 2MASS catalog (Strutskie et al. 2006) and therefore
all have (J--K) available.  The aim is to focus on utilizing near-IR data and then
compare to published literature results.

All of the red giants here have parallaxes measured by Hipparcos and thus have distances
known to various levels of accuracy.  The parallax is used to set the distance and, thus,
M$_{\rm K}$.  Bolometric corrections from Bessell, Castelli, \& Plez (1998) are used in conjunction
with (J--K) to determine M$_{\rm bol}$ and consequently stellar luminosity.  Basic data
for the red giants are presented in Table 1.

The most important derived quantity affecting the overall shape of the spectrum is T$_{\rm eff}$, which here
is based on an average of two T$_{\rm eff}$-(J--K) calibrations; one from Gonzalez Hernandez
\& Bonifacio (2009) and the other from Bessell et al. (1998).  The Gonzalez Hernandez \&
Bonifacio calibration is defined in the 2MASS photometric system of (J--K$_{\rm S}$), so the
original Johonson (1965) (J--K) colors and K magitudes for these bright red giants (which have saturated
magnitudes in the 2MASS catalog) were transformed
to 2MASS values using the prescription from Carpenter (2001).  The differences between the two calibrations,
for a given metallicity, are less than 30-50K for T$_{\rm eff}$$\le$4500K, but above this temperature
the Bessell et al. (1998) becomes cooler by about 100K for a given (J--K$_{\rm S}$) (where we have
transformed the Bessell et al. scale onto the 2MASS system).  There is also a small metallicity effect on the
(J--K$_{\rm S}$) calibration which, for a given (J--K$_{\rm S}$) can be $\sim$100K over the broad metallicity
range of [m/H]= -2.0 to +0.0 (where m is taken to represent the generic abundance of metals, i.e. those elements
other than H and He).  Given differences between two independent T$_{\rm eff}$ calibrations of less than
30-50K over the temperature range of the target stars, as well as their modest range in metallicity, and 
well-determined magnitudes with $\sim$ 0.02 uncertainty in (J--K$_{\rm S}$) (which translates to 
$\Delta$T$_{\rm eff}$=30K), the derived values of T$_{\rm eff}$ have uncertainties of $\pm$50K or less. 

Given the stellar luminosity and effective temperature, these fundamental properties
are then compared to stellar evolution models presented in Bertelli et al. (2008; 2009, with
data from these papers taken from the website http://stev.oapd.inaf.it/YZVAR )
to estimate stellar masses and, ultimately, surface gravities through:

g/g$_{\odot}$= (M/M$_{\odot}$) $\times$ (L$_{\odot}$/L) $\times$ (T$_{\rm eff}$/T$_{\rm eff \odot}$)$^{4}$.


Determining the surface gravity from evolutionay tracks is an iterative process,
because it requires knowledge of the stellar metallicity.  This is illustrated in
Figure 2 where the stellar luminosity and T$_{\rm eff}$ are plotted along with evolutionary
models from a range of masses, with each panel having a different metallicity: Z=0.017
(top panel), 0.008 (middle panel), and 0.004 (bottom panel), correpsonding to overall
metallicities of [m/H]= +0.07, -0.26, and -0.56, respectively, when taking Z=0.0145 for the Sun, as suggested by
Lodders (2010).  The illustrated evolutionary tracks follow stellar evolution up to the tip of
the first ascent red giant branch (which we label RGB).  
Initial model metalliities for the stars studied here are based on previously published results, with $\beta$ And,
$\delta$ Oph, and HD199799 started with solar metallicity, $\alpha$ Boo with [Fe/H]= -0.5, and $\mu$ Leo
with [Fe/H]= +0.3.  As there are numerous Fe I lines in this spectral region, a sample of Fe I lines are used
to determine the iron abundance.  Given the limited range of metallicity spanned by the nearby red giants studied
here, the Fe abundance is a good proxy for the overall stellar metallicity.  If the derived Fe abundance differed
by more than 0.1 dex from the assumed value, a new surface gravity was derived and a new model generated
in order to reanalyze the Fe I lines until convergence between model abundance and derived abundance.  For
these relatively well-studied red giants, no more than two iterations were required until convergence.
Table 2 presents, for these red giants, the derived parameters T$_{\rm eff}$, Log (L/L$_{\odot}$), mass, log g,
microturbulent velocity ($\xi$ -- see Section 4.2), and overall metallicity, which is represented
by [Fe/H] as discussed in Section 4.2. 

\subsection{Stellar Parameters and Evolutionary State}

Given the position of the stars in the Log L versus T$_{\rm eff}$ plane shown in Figure 2,
it can be ascertained that 4 stars ($\alpha$ Boo, $\mu$ Leo, $\delta$ Oph, and
$\beta$ And) are most likely first-ascent red giants of low to moderate mass.  Based on its
warmer temperature and somewhat lower luminoisty, it may be more likely 
that $\mu$ Leo is in, or very near the core-He burning phase (clump giant) and has thus already
ascended the RGB and experienced the core-He flash.  It is also possible that the other 3 stars
already noted above may
be post-He core-burning stars ascending the AGB; however, lifetimes on the two separate 
evolutionary sequences would favor them as being on the RGB.  All fall above the ``luminosity
bump'' (Fusi Pecci et al. 1990) and would have thus experienced any extra-mixing that may occur 
during that phase of stellar evolution.

The coolest and most luminous red giant in this sample, HD199799, is an AGB star that is
experiencing the third dredge-up, as evidenced by the observation of Tc I in its spectrum
and its elevated s-process abundances, such as [Y/Fe]$\sim$+0.5 or [Nd/Fe]$\sim$+0.4
 (Smith \& Lambert 1988; 1990).

\section{The Abundance Analysis}

As with the initial abundances to be derived from APOGEE, the analysis here will rely on
1-D model atmospheres computed in LTE.  Within APOGEE, the choice was made to base
the initial abundance determinations on an extensive model atmosphere database generated
using the  ATLAS9 code (Kurucz 1993), thus the model atmospheres used here are based
on these APOGEE models.  ATLAS9 is widely used as a universal LTE 1−D plane-parallel 
atmosphere modeling code.  The details of constructing the atmosphere database for APOGEE
using ATLAS9 are presented in Meszaros et al. (2012).

The individual elemental abundances themselves
will also be derived under the assumption of LTE with the spectrum synthesis code MOOG
as the analysis tool (Sneden 1973 is the original reference, but updated information, data,
and the most recent versions of the code can be found at 
http://www.as.utexas.edu/~chris/moog.html).

An extensive effort was made to produce as complete a spectral linelist as possible, so the
analysis technique used here, as in APOGEE in general, is to utilize spectrum synthesis in
a quantitative comparison between synthetic and observed spectra.  All abundances are determined
for each spectral line, or group of closely spaced molecular lines, via minimizing the residuals
between observed and synthetic spectra as a function of the abundance in question.

\subsection{Choice of Fe I Lines and the Microturbulence}

Due to both its relatively large cosmic abundance and large number of energy
levels, iron has historically been used as a diagnostic species for certain
stellar parameters, as well as an overall metallicity indicator, so the results for
Fe I lines selected from the APOGEE linelist (Shetrone et al. 2012) are discussed
here.  In particular, Fe I lines are well-suited for setting the microturbulent velocity, $\xi$, that is
required for 1-D model atmosphere abundance analyses.  

A large number of Fe I lines fall within the APOGEE spectral window 
(15100 -- 16900 \AA), although there are no Fe II lines strong enough to be detected
in these red giants.  Since all abundance determinations are done via spectrum
synthesis, the Fe I lines were culled to include those lines which contain
only contributions from Fe I (this was determined by preliminary synthesis of the red
giant spectra, then setting the Fe abundance to
zero and ensuring that the spectral feature in question vanished); since the Fe I lines
are so numerous, this stringent selection still results in an adequate list of lines.
Table 3 presents the Fe I lines used for the iron abundance determinations.  As is
true of most of the atomic lines in the H-band, the excitation energies are dominated
by rather high-excitation lines, although for Fe I there are two lower-excitation
lines.  

The Fe I lines span a large enough range in line-strength that they can be used
to set the microturbulent velocity.  This parameter is defined by the value of $\xi$
that produces no trend in the Fe abundance as a function of line-strength.  For a
range of microturbulent velocities Fe abundances were determined from each Fe I line
and the adopted value of $\xi$ was one in which a linear regression between A(Fe)
as a function of log(W$_{\lambda}$/$\lambda$) yielded a slope of zero.  This procedure
is illustrated graphically in Figure 3, where the abundances derived for each individual
line is plotted versus $\xi$ (the continuous lines); the circle with errorbars shows
where the slope of A(Fe) with line-strength goes to zero.  

The microturbulent velocities set by the Fe I lines, as well as the Fe abundances are listed
in Table 2 as part of the derived set of stellar parameters.  The individual line-by-line
abundances as determined for each Fe I line are also listed in Table 3.  The Fe abundances were
used to set the final overall metallicities of the ATLAS9 model atmospheres for the analysis of
additional elements via atomic or molecular lines.

\subsection{The C, N, and O Abundances}

Abundances for $^{12}$C, $^{14}$N, $^{16}$O, and the minor carbon isotope $^{13}$C are
derived from combinations of vibration-rotation (V--R) lines of CO (X$^{1}$$\Sigma$$^{+}$)
and OH (X$^{2}$$\Pi$), along with electronic transitions of CN (A$^{2}$$\Pi$ -- X$^{2}$$\Sigma$).
Although the details of the ingredients that went into these molecular lines in the APOGEE
linelist will be found in Shetrone et al. (2012 -- in preparation), a few highlights are noted
here.  The adopted dissociation energies (D$_{0}$) are 11.092 eV for CO, 4.395 eV for OH, and
7.76 eV for CN.  The gf-values are from Goorvitch (1994) for CO, Goldman et al. (1998) for
OH, and the CN gf-values are taken from the baseline linelist of Kurucz (1993), with updates to
these values from the prescription given in Melendez \& Barbuy (1999).

The procedure used for the CNO analysis, since all red giants studied here have C/O $\le$ 1, is
to use CO to set the carbon abundance.  With that carbon abundance, OH provides an O-abundance,
which, if different from the initial value (taken to be the solar value scaled by the stellar value of [Fe/H]),
is used to re-analyze the CO lines.  This process is
repeated until the C and O abundances yield consistent abundances from CO and OH.  With
these values of C and O, the CN lines are used to derive the abundance of nitrogen.  In general,
the abundance of N has little to no effect on the CO and OH lines, however, the respective final
C, N, and O abundances all provide self-consistent results from CO, OH, and CN.

All spectral features were synthesized and the molecular regions that were used to extract the
abundances are listed in Table 4.  In this case, if a range of wavelengths is given it indicates
that a spectral window containing numerous molecular lines was used (for CO and OH), while
for CN individual transitions were identified and synthesized.  As with Table 3, the abundances
derived for each feature are also listed in Table 4.  In the case of $^{13}$C, as has been customery
for minor isotopes, the abundance is listed as the ratio of $^{12}$C/$^{13}$C, with A($^{12}$C) being
set by the mean abundance of carbon-12.

\subsection{The Other Atomic Lines and Elemental Abundances}

With the Fe and C, N, and O abundances in hand, additional elements are derived via atomic spectral
lines in the wavelength interval 15100 -- 16900\AA.  The APOGEE linelist was searched for
suitable species, all neutral over the T$_{\rm eff}$ range of the red giants here, and some 12 other
elements were deemed detectable and able to be analyzed quantitatively: Mg I, Al I, Si I, K I, Ca I, Ti I,
V I, Cr I, Mn I, Co I, Ni I, and Cu I.  The additional atomic lines are listed in Table 5.  As with the molecular
lines, if a precise wavelength is shown, the transition in question is represented by a single line, however
if the wavelength is approximated by an integer value it indicates multiple transitions due to hyperfine (hfs)
splitting or isotopic splitting.  In the case of isotopic components, solar-system isotopic fractions are
assumed and, in general, this assumption has no significant effect on the derived total atomic abundances.
Figure 4 illustrates the fitting procedure for an atomic line from Mn I in $\delta$ Oph, with the underlying
hfs components illustrated, as well as nearby blending features, with several synthetic spectra plotted 
which have differing Mn abundances.
As with the previous Tables 3 and 4, individual abundances are provided for each line or feature.  As noted
previously, the details of the generation of the APOGEE linelist will be found in Shetrone et al. (2012).

Table 6 provides a summary table of all abundances listed as their mean values and standard deviations
from the lines or spectral intervals shown in Tables 3, 4, and 5.  In cases where only one line or region
was observed, no standard deviation is given.  The abundance of $^{13}$C is given as the isotopic ratio of 
$^{12}$C/$^{13}$C.

\subsection{Abundance Sensitivity to Stellar Parameters}

Most of the abundances presented here are derived from more than one atomic or molecular line or features
of lines, thus internal consistency in the average abundances are defined by the scatter from the individual
features.  This scatter is characterized by the standard deviations of Table 6.  Because multiple features
arise from differing line strengths (weak versus strong), excitation potentials, or ionization fractions, these
respective standard deviations provide some estimate as to how well the 1-D LTE analysis can recover the
details of the underlying stellar spectrum from the basic model atmosphere parameters T$_{\rm eff}$, log g,
metallicity, and derived value of $\xi$.  The typical standard deviations from Table 6 are $\sim$$\pm$0.05 dex.
Of note, is the metal-rich K-giant $\mu$ Leo, which exhibits exceptionally strong CN lines, that are numerous
throughout this H-band window.  The background CN absorption, creates blending which lowers the overall
signature of other abundances and increases the uncertainty in these other abundances; this is reflected in
the generally larger standard deviations found in the $\mu$ Leo abundances.

Besides internal consistency from the various spectral lines of different species, it is useful to explore their
sensitivities to stellar parameters in order to characterize the magnitudes of possible systematic effects due to
uncertainties in the four primary parameters of effective temeprature, surface gravity, metallicity, and 1-D derived 
microturbulent velocity. For this discussion T$_{\rm eff}$ is written as T, log g as G, model atmosphere metallicity
([m/H]) as m, and microturbulent velocity as $\xi$.  Because the abundance for a given element, 
A=log[N(A)/N(H)] + 12. = A(T,G,m,$\xi$), the incremental change in this abundance due to small
variations in the parameters is given by  

dA=($\partial{A}/\partial{T}$)dT + ($\partial{A}/\partial{G}$)dG + ($\partial{A}/\partial{\xi}$)d$\xi$ +
($\partial{A}/\partial{m}$)dm.

The scatter in the abundance caused by small changes in the stellar parameters is then defined as

$\sigma$$^{2}$$_{\rm A}$=$\langle$dA$^{2}$$\rangle$ 

and dA$^{2}$ can be written as

(dA)$^{2}$=[($\partial{A}/\partial{T}$)dT]$^{2}$ + [($\partial{A}/\partial{G}$)dG]$^{2}$ + 
[($\partial{A}/\partial{\xi}$)d$\xi$]$^{2}$ + [($\partial{A}/\partial{m}$)dm]$^{2}$

+($\partial{A}/\partial{T}$)($\partial{A}/\partial{G}$)[dTdG + dGdT]

+($\partial{A}/\partial{T}$)($\partial{A}/\partial{\xi}$)[dTd$\xi$ + d$\xi$dT]

+($\partial{A}/\partial{T}$)($\partial{A}/\partial{m}$)[dTdm + dmdT]

+($\partial{A}/\partial{G}$)($\partial{A}/\partial{\xi}$)[dGd$\xi$ + d$\xi$dG]

+($\partial{A}/\partial{G}$)($\partial{A}/\partial{m}$)[dGdm + dmdG]

+($\partial{A}/\partial{\xi}$)($\partial{A}/\partial{m}$)[d$\xi$dm + dmd$\xi$].

The partial derivative terms are calculated by changing each model parameter and determining the
resulting change in logarithmic abundance, A.  Since the perturbation in stellar parameters is relatively
small, the changes are approximated by linear trends.  As this paper is a limited study of a few stars
as a test of the APOGEE linelist, a test of abundance sensitivity to stellar parameters is carried out for
a representative model taken with T$_{\rm eff}$=4000K, log g=1.3, [m/H]=+0.0, with $\xi$=2.0 km-s$^{-1}$.
Predicted equivalent widths were generated for the lines used in the abundance analysis and the stellar parameters
were then pertrubed by dT=+50K, dG=+0.2 dex, dm=+0.1 dex, and d$\xi$=+0.2 km-s$^{-1}$, with new
abundances derived for each separate perturbation, giving the change in abundance for each parameter change.
These coefficients, which reflect the change in abundance, are listed in Table 7 and provide an overall view of how 
sensitive the derived abundances are to changes in stellar parameters that are representative of realistic errors for these 
particular red giants.

Columns 6 and 7 in Table 7 provide the values of dA two cases: one using only the first four terms which contains the
change in abundance due to changes in each primary parameter (dA' in column 6, which is valid if all parameters are
independent of each other), while column 7 includes
the covariant terms.  Within the analysis technique applied here, it was found that the microturbulent velocity did not
measurably depend on the
derived effective temperature or gravity, and the dT-d$\xi$ and dG-d$\xi$ cross terms are not included.  As can be
seen by the comparison of dA' and dA, the covariant terms contribute significantly, in many species, to the uncertainty. 
The largest covariant terms are typically those between T$_{\rm eff}$ and log g, caused by the slope of the RGB in the
T$_{\rm eff}$--log(L/L$_{\odot}$) plane (since L$\sim$T$_{\rm eff}^4$/g).

\subsection{Comparison to Previously Published Abundances}

The abundances derived here rely on the APOGEE H-band spectral window from $\lambda$15100 to
16900\AA, along with parallaxes and photometry.  Historically, most detailed abundance patterns obtained 
from high-resolution spectroscopy have
tended to be based on wavelengths below $\lambda$10000\AA\ so it is useful to compare the abundances
based on the APOGEE linelist with those from other studies for stars in common.  Within the sample of red giants
discussed here, previous high-resolution abundance analyses have been conducted on $\alpha$ Boo,
$\mu$ Leo, $\beta$ And, and HD 199799.

\subsubsection{$\alpha$ Boo}

Being the nearest red giant, $\alpha$ Boo is often used as the standard star of choice for abundance
comparisons.  Ramirez \& Allende Prieto (2012) have produced a recent and detailed high-resolution
optical spectral analysis, revieiwng both fundamental stellar parameters for this star, as well as
deriving an extensive set of elemental abundances.   They used the observed spectral energy distribution
and compared to theoretical ones from the Kurucz grid of no-overshoot model atmospheres with
$\alpha$-element enhanced compositions ([$\alpha$/Fe]=+0.4; Castelli \& Kurucz 2003) in order to
set the T$_{\rm eff}$ for $\alpha$ Boo.  The flux from $\lambda$ $\sim$0.3 -- 10 $\mu$m was used
with the resultant best-fit T$_{\rm eff}$=4286$\pm$30K.  This value is entirely consistent with that
obtained here (T$_{\rm eff}$=4275$\pm$50K) based on the (J--K) color and calibration used here from Bessell et al.
(1998) and Gonzalez Hernandez \& Bonifacio (2009).

Ramirez \& Allende Prieto (2012) used stellar evolutionary model tracks to estimate the mass (as well as
age) of $\alpha$ Boo, as the effective temperature and parallax are known to high precision.  The mass was
set by comparison to Yonsei-Yale isochrones (Yi et al. 2001; Kim et al. 2002) and was found to be 
M=1.08$\pm$0.06M$_{\odot}$, which yields a surface gravity of log g=1.66$\pm$0.05.  In this study the
T$_{\rm eff}$  =4275K value from the (J--K) color is adopted, while the Ramirez \& Allende Prieto log g is used.
These stellar parameters were used to generate a model atmosphere as described in Section 4 with the ATLAS9
code in the same manner as is being done for the APOGEE targets (Meszaros et al. 2012).

The microturbulence, $\xi$, derived by Ramiriz \& Allende Prieto was set by Fe I lines and was found to be
$\xi$=1.74 km-s$^{-1}$.  The Fe I lines used from the APOGEE linelist yielded a value of $\xi$=1.85$\pm$0.05
km-s$^{-1}$: very close to that derived from optical Fe I lines.  

A direct comparison of the derived abundances here with those from Ramirez \& Allende Prieto (2012) using
completely differsent sets of lines leads to the following mean difference and standard deviation for 13 elements
in common of $\Delta$A(x)(This paper -- Ramirez \& Allende Prieto)=+0.03$\pm$0.10 dex.  There is no significant
offset in the overall abundance patterns and the scatter is that expected given the uncertainties in the
log gf scales and internal line-to-line scatter within each study.  This comparison suggests that abundances
derived from the APOGEE linelist can be used for comparisons to those found from analyses at visual wavelengths
when high-quality spectra are analyzed.

\subsubsection{$\mu$ Leo}

This K-giant is often cited as the ``prototype'' for metal-rich stars, having been noted as such by Spinrad
\& Taylor (1969), and some of its spectral
regions contain numerous and strong CN lines.  Two older abundance studies in the literature
(Gratton \& Sneden 1990; Smith \& Ruck 2000) are used to compare to the abundances derived here, as well as a
more recent analysis by Fulbright, McWilliam, \& Rich (2007--FMR).  As with
the $\alpha$ Boo comparison in Section 4.5.1, both of these published analyses are based on high-resolution
spectra in the visual.  Both studies used T$_{\rm eff}$=4540K for $\mu$ Leo, which is similar to the value 4550K
adopted here.  The surface gravities were slightly different, with log g= 2.3$\pm$0.3 in Gratton \& Sneden and
2.2$\pm$0.1 in Smith \& Ruck; both values agree with the value here, within the uncertainties, of log g=2.1 derived 
from the luminosity and estimate of stellar mass. 

The value of $\xi$=1.8 km-s$^{-1}$ found from the H-band Fe I lines is slightly higher than 1.2 km-s$^{1}$
found by both Gratton \& Sneden (1990) and Smith \& Ruck (2000), but with all studies having uncertainties of
0.3-0.5 km-s$^{-1}$, this offset cannot be claimed to be significant.  A comparison of abundances between
here and Gratton \& Sneden (1990) for 12 elements finds a mean $\Delta$A(x)(This paper - GS)=-0.05$\pm$0.14 dex, 
while the same comparsion with Smith \& Ruck (2000) yields $\Delta$A(x)=-0.05$\pm$0.02 dex for the 3 elements
analyzed by them: Fe, Mg, and Ca.  The abundance offsets are not significant and the scatter is within what
is expected.

The more recent abundance analysis of $\mu$ Leo by FMR was part of their study of bulge red giants.
Their derived stellar parameters were quite similar to those found here, with 
T$_{\rm eff}$(FMR)=4520K, ($\Delta$T(this study - FMR)= +30K),
Log g = 2.33 ($\Delta$log g(This study - FMR)= -0.23), and
$\xi$=1.50 km-s$^{-1}$ ($\Delta$$\xi$(This study - FMR)= +0.3 km-s$^{-1}$).
There are 7 elements in common (Fe, O, Mg, Al, Si, Ca and Ti)
and the mean difference is $\Delta$A(x)(This study - FMR)=-0.10 $\pm$ 0.13 dex.
A comparison amongst the three $\mu$ Leo studies points to rather small
offsets in stellar parameters ($\sim$ 30K in T$_{\rm eff}$, 0.2 dex in log g,
0.3-0.5 km s$^{-}1$ in $\xi$, and 0.1 dex in [m/H]) with the conclusion
that H-band spectroscopy can be compared reliably to optical
analyses with reasonable accuracies, even at this higher metallicity.

\subsubsection{$\beta$ And}

This M-giant was analyzed previously by Smith \& Lambert (1985) who derived T$_{\rm eff}$=3800K (from its
(V--K) color), log g=1.6, and $\xi$=2.1 km-s$^{-1}$.  The derived effective temperature here of 3825K is
very close to Smith \& Lambert, as is the microturbulence of 2.2 km-s$^{-1}$.  The surface gravity from
Smith \& Lambert is larger by 0.7 dex, but their value relied on the distance (as it does in this paper), for
which they used a luminosity calibration for the Ca II K-line reversal that was tied to a pre-Hipparcos
Hyades distance that was too small.  Making the estimated log g corrections to the Smith \& Lambert (1985)
abundances (based on their published sensitivities of abundance with log g) for $^{12}$C, $^{14}$N, $^{16}$O, Ti, Fe, and Ni we find a mean difference and standard
deviation of $\Delta$A(x) (This paper - Smith \& Lambert)=+0.02$\pm$0.14 dex. 

Beta And was also used as a comparison star by Chou et al. (2010) in their study of Sgr dwarf-galaxy tidal
streams in the Galactic halo.  Chou et al. (2010) applied a similar technique as that employed here to
estimate temperatures and gravities and found T$_{\rm eff}$=3850$\pm$75K, log g=0.9 and $\xi$=1.96
km-s$^{-1}$, giving differences of only 25K in T$_{\rm eff}$, 0.0 in log g, and +0.24 km-s$^{-1}$ in $\xi$.
Their analysis used the spectral region around $\lambda$7500\AA\ containing a sample of Fe I and Ti I lines and
determined A(Fe)=7.12$\pm$0.06 and A(Ti)=4.55$\pm$0.12, which can be compared to values of 7.23$\pm$0.03
and 4.72$\pm$0.02, respectively, here, yielding differences of +0.11 dex in Fe and +0.17 dex in Ti.  

\subsubsection{ HD 199799}

HD 199799 is classified as spectral type M2S, which means that it displayed enhanced bands of ZrO in
classification spectra.  Smith \& Lambert (1988) detected the radioactive s-process element Tc I in its
spectrum, identifying this star as an intrinsic TP-AGB star undergoing third dredge-up.  A quantitative
abundance analysis was then carried out by Smith \& Lambert (1990) and they derived similar stellar
parameters to those derived here: the same T$_{\rm eff}$=3400K, with a slightly lower value of log g=+0.3,
compared to +0.5 here.   The microturbulent velocity determined here from the H-band spectra is 
$\xi$=2.4 km-s$^{-1}$, which is 0.7 km-s$^{-1}$ larger than Smith \& Lambert (1990), who used Fe I lines near 
$\lambda$7500\AA.   Taking into account the slightly lower gravity and smaller $\xi$, the abundances
from Smith \& Lambert (1990) can be compared to those here for Fe, $^{12}$C, $^{14}$N, $^{16}$O, Ti, Cr,
Co, and Ni, with the mean difference being $\Delta$A(x)( Us - SL90)= -0.08$\pm$0.13.   The respective carbon
isotopic ratios compare well, with Smith \& Lambert (1990) finding $^{12}$C/$^{13}$C=28, while a value of 27 is
obtained here.  The derived C/O ratios are the same in the two studies, with $^{12}$C/$^{16}$O=0.68. 

The comparisons presented here are between studies that have possibly used somewhat different families of model
atmospheres (e.g. MARCS or ATLAS), different spectral regions, and linelists derived from heterogeneous sources.
Given the variety of data and analysis techniques, the derived abundances, in all cases, compare well
at approximately the 0.10 to 0.15 dex level.  A glance at Table 7 reveals that such differences fall within this
range of uncertainty for most elements given modest changes in fundamental model atmosphere parameters.  Since
the derived stellar parameters themselves rely to some degree on the type of analysis, the generally good
agreement between the derived abundances from the various published sources, along with the expected uncertainties
in this study, as presented in Table 7, is encouraging.   The gist of the comparisons, since many of the studies are
based on optical spectra, is that H-band spectrscopy can be used to define a precise internal abundance 
scale that can also extended to abundances derived from other, more common spectral windows, such as in the
optical.

\section{Discussion}

The abundances from Table 6 and an estimation of their respective uncertainties from Table 7, along with the 
stellar parameters from Table 2 represent, to some extent, what ASPCAP will produce from the APOGEE spectra.  
The following discussion highlights these abundances in light of internal red giant stellar evolution, as well as Galactic
chemical evolution and stellar populations. These discussion topics anticipate what sorts of information will be gleaned 
from the APOGEE targets. 

\subsection{CNO Abundances and Red-Giant Evolution}

Based on their luminosities and effective temperatures (Figure 2), four of the red giants studied here are either
first-ascent giants or, perhaps for $\mu$ Leo, a clump red giant.  HD 199799 is in a more advanced stage of
stellar evolution where it is undergoing thermal pulses and third
dredge-up on the AGB.  The combination of nucleosynthesis and convective mixing in evolved red
giants should result in surface abundances of $^{12}$C, $^{13}$C, and $^{14}$N that have been altered
from their initial (main sequence) values due to CN-cycle H-burning and first dredge-up.

The result of CN-cycle mixing is seen mostly easily in the $^{12}$C/$^{13}$C ratios which are much higher
in main-sequence stars, with the Sun having a value of 89.  All four of the RGB or clump giants have quite
low values, ranging from $^{12}$C/$^{13}$C$\sim$6-20.  Gilroy (1989) and many subsequent studies of 
both open and globular clusters (e.g., Mikolatis et al. 2012 or Briley et al. 1995) have
noted that the carbon isotopic ratios correlate with stellar mass in stars above the luminosity bump on
the RGB or in the clump; red giants more massive than M$\sim$2.5M$_{\odot}$ have values of 
$^{12}$C/$^{13}$C$\sim$22-30, while lower-mass red giants exhibit decreasing isotopic ratios with
declining mass.  The mass estimates derived here for field stars from distance-luminosity and stellar evolutionary
tracks are modestly well-defined and the top panel of Figure 5 shows the $^{12}$C/$^{13}$C values versus the 
estimated stellar mass.  Also plotted are representative results for open clusters from Gilroy (1989) and
Mikolaitis et al. (2012), and the globular cluster M71 from Briley et al. (1995).  The cluster red giants have
recently evolved from the cluster turn-offs, which have a well-determined masses.  For the lower-mass stars
included in these cluster studies, evolution along the RGB is rapid ($\sim$10$^{7}$ yrs) when compared to
main sequence lifetimes ($\sim$10$^{8}$-10$^{9}$ yrs) so the red giant mass is similar to the turn-off mass.
This can be demonstrated using the evolutionary tracks from Bertelli et al. (2009), where an isochrone with a turn-off mass
of 4.0M$_{\odot}$ has a red giant tip mass of 4.2M$_{\odot}$, or a ratio of RGB to TO masses of 1.05.  
This ratio of masses becomes even closer to 1.0 for lower masses, which include all of the points in Figure 5.
The points from Gilroy 
represent the mean values of $^{12}$C/$^{13}$C for 19 open clusters with a total of 55 red giants in her sample.  
The mean carbon isotopic ratios for two other open clusters (with a total of 10 red giants from Cr261 or NGC6253) 
from Mikolaitis et al. (2012) are also shown (open blue squares). 
A representative globular cluster, M71, which has
a lower main-sequence mass (M$\sim$0.9M$_{\odot}$) from Briley et al. (1995) is plotted as the open blue
triangle and represents the mean of two of the CN-weak stars in this cluster.  The CN-strong giants that
Briley et al. also studied are not plotted, as they exhibit slightly lower isotopic ratios but almost certainly
represent 2nd generation stars in M71 that were formed from large fractions of the ejecta of 1st generation
intermediate-mass red giants and thus are not representative of single-star red giant evolution.  

The curves in the top panel of Figure 5
represent stellar evolutionary model results from Charbonnel \& Lagarde (2010); the solid black
line illustrates their predictions for standard RGB dredge-up (no ``extra-mixing'' or rotation included). 
Clearly, standard RGB first dredge-up does not fit the observed trend of $^{12}$C/$^{13}$C for red giants
with M$\le$2.2M$_{\odot}$.  This poor comparison between theory and observation has led to work in 
trying to identify other types of ``extra mixing'' 
mechanisms that would lead to lower carbon-isotopic ratios in the lower-mass red giants.
Two such hypothesized mechanisms are thermohaline mixing or thermohaline mixing coupled with
rotational mixing, as discussed recently by Charbonnel \& Lagarde (2010).  The dashed curves
present results from Charbonnel \& Lagarde also, with the short dashes for thermohaline mixing only 
and the long dashes for thermohaline mixing plus rotational mixing for models with initial equatorial
rotational velocities of 110 km-s$^{-1}$.   The stellar models that incorporate these particular types of
extra-mixing provide fair agreement with the observations, although these are not the only types of
processes that have been studied and are used here to merely illustrate the effect of extra mixing.  The
trends found in the clusters also overlap well with the field stars analyzed here, except for HD 199799,
which is in a different evolutionary phase than the other red giants.  This star has enriched its surface in $^{12}$C
that was synthesized by He-burning as a TP-AGB star.  In the simplest case of pure $^{12}$C dredge-up, 
the surface $^{12}$C/$^{13}$C ratio would
increase, which is where HD 199799 lies in the top panel of Figure 5 relative to the other red giants.  
Such an effect has been noted by both
Lambert et al. (1986) and Smith \& Lambert (1990), where these studies find increasing values of $^{12}$C/$^{13}$C
as TP-AGB stars evolve from M to MS to S to C stars (with increasing values of $^{12}$C/$^{16}$O).

The bottom panel of Figure 5 plots the abundance by number ratio of N(C)/N(N) versus stellar mass, where
the carbon abundance is now the total of $^{12}$C + $^{13}$C: for reference the solar ratio is 3.80.  Nitrogen
is also the total abundance, however in the case of the observations, $^{15}$N is ignored as its abundance is very low
(and not detected in these red giants) when compared to $^{14}$N.  There is
an observed decrease in C/N with increasing mass as predicted by both standard RGB dredge-up and
dredge-up plus thermohaline mixing (Charbonnel \& Lagarde 2010), which are shown as the solid and dashed
curves, respectively, and are from the same models as in the top panel.   
In the case of the C/N ratios, the total carbon and nitrogen abundances
are not as strongly influenced by non-standard mixing as the carbon isotopic ratios. 

Figure 6 combines the C/N ratios versus $^{12}$C/$^{13}$C ratios for both the observed red giants and the
models from Charbonnel \& Lagarde (2010).  When displayed this way, the different behavior between
C/N versus $^{12}$C/$^{13}$C is striking and these two values are purely spectroscopically derived
quantities.  The trend of C/N with $^{12}$C/$^{13}$ seems to provide a sensitive indicator of red giant
mixing processes, as well as yields a mass estimate for stars on the RGB.  Again, HD 199799 stands apart
from the other red giants as it has internally enhanced its $^{12}$C abundance by about a factor of two since
it was a RGB star, based on its $^{12}$C/$^{13}$C and mass, relative to other RGB stars in the top panel of 
Figure 5.  The dashed-dotted line connects to the region in this diagram where HD 199799 would fall if its
surface $^{12}$C abundance were reduced by a factor of two, which is a simplified sketch of how TP-AGB
stars might evolve across such a diagram from the RGB.

The assumption of only CN-cycle mixing taking place in the red giants is tested in Figure 7, where
the total carbon abundance ($^{12}$C + $^{13}$C) is plotted versus the nitrogen abundance (here $^{14}$N).
The solar values of C and N are also shown, along with the dashed line illustrating scaled solar-abundances
for both C and N.  The continuous blue curves represent decreasing carbon and increasing nitrogen
abundances from the scaled solar lines, with the constraint that the number of C + N nuclei is
conserved, as in the CN-cycle (with no leakage into the ON-cycle).  Each curve is scaled in its initial
 C and N abundances by the stellar value of [Fe/H] for $\alpha$ Boo, $\beta$ And, $\delta$ Oph, and
$\mu$ Leo.  CN-cycle mixing alone is sufficient to represent the derived abundances of C and N
with initial abundances having [C + N/Fe]$\sim$0.0 for each RGB star.  
Due to having undergone third dredge-up and $^{12}$C enrichment, HD 199799
will be offset from its CN curve, so no curve is plotted for it.  The dashed-dotted line simply shows the
shift in A(C) if HD 199799 has doubled its RGB $^{12}$C abundance.  Having nearly the same metallicity as
$\beta$ And, HD 199799 might be expected to have been near the CN curve for $\beta$ And before it
began its evolution as a TP-AGB star.

The APOGEE H-band wavelength region presents good opportunities for probing many nuclear and
mixing processes that occur over the various phases of red giant evolution.

\subsection{Element Ratios and Galactic Chemical Evolution}

Other than the C and N isotopes, the vast majority of APOGEE red giant targets will not have altered
their surface abundances of the other studied elements, so the remaining elements are dominated by 
chemical evolution within the various stellar populations of the Milky Way (or its accreted systems).  
The derived abundances of each of the other elements are discussed in the light of observed trends
of these elements with A(Fe) from other studies in the literature.  The number of published abundances
is now enormous; however in this first exploratory study using the APOGEE linelist, a comparison
with a small number of selected literature sources is sufficient.   For comparison purposes in this
paper a stellar sample representative of the Milky Way stellar populations was constructed using
Reddy et al.  (2003; thin disk with some thick disk and having 58 stars), Reddy et al. (2006; thick disk
with 176 stars), Johnson (2002; halo with 23 stars), and Fulbright (2002; halo with 178 stars).

The investigation of elements derived from the APOGEE linelist is via a number of figures, beginning with
Figure 8, which presents values of O/Fe, Mg/Fe, and Al/Fe versus A(Fe); the shorthand notation here of x/Fe
denotes the logarithmic number ratio of element x to iron, log[N(x)/N(Fe)], or A(x) - A(Fe).  This nomenclature
is chosen over the [x/Fe] in the plots as it presents a direct logarithmic number ratio with the solar abundance 
ratio being noted in each panel, thus no reference to
the underlying solar abundance is needed, but the relative scales are equivalent to [x/Fe].  The top panel
of Figure 8 shows the behavior of O/Fe and, as has been well-known for decades, values of O/Fe are
elevated in the thick disk and halo stars.  Among the field red giants here, the slightly metal-poor
objects $\alpha$ Boo and $\beta$ And both have O/Fe enhanced by $\sim$+0.2 -- +0.3.  The near-solar 
metallicity M-giant
$\delta$ Oph has a near-solar ratio of O/Fe, while the metal-rich giant $\mu$ Leo also exhibits a
near-solar value O/Fe.

The behavior of Mg/Fe is shown in the middle panel of Figure 8 and, like oxygen, Mg/Fe ratios are
typically elevated in the thick disk and halo populations.  The four red giants closest to solar in A(Fe) all
exhibit basically solar Mg/Fe ratios, with $\alpha$ Boo having a slight enhancement of$\sim$+0.1
dex, which overlaps the thick disk populations.

Aluminum abundances are plotted in the bottom panel of Figure 8 and the behavior of Al/Fe versus A(Fe)
is not so well-defined, with relatively large scatter.  Mu Leo is found to be modestly enhanced in Al/Fe by
about +0.10--+0.15 dex, as is $\alpha$ Boo, with an enhancement of $\sim$+0.2.  

The abundances for the rest of the elements studied here are illustrated in Figure 9 (for Si, K, and Ca), Figure 10
(for Ti, V, and Cr), Figure 11 (for Mn, Co, and Ni), and Figure 12 (for Cu).  In all elements, the general trends found
in the general Galactic populations are recovered in the small sample of field red giants included here.  The one
exception is Ti in $\delta$ Oph and $\mu$ Leo, both of which are found to exhibit enhanced values of Ti/Fe of
+0.2 dex.  There is no obvious explanation for this, however the continuing analysis of APOGEE spectra with this
linelist wuill reveal if this possible effect appears in significant numbers of other red giants.  The results presented
and discussed here indicate that H-band spectra analyzed in LTE with the APOGEE linelist and 1D model atmospheres
can provide accurate results for chemical abundances.

\section{Conclusions}

Archival high-resolution FTS spectra from $\lambda$15100--16900\AA\ are analyzed to derive detailed
chemical abundance distributions for five field red giants using the APOGEE linelist and model atmosphere
grid.  The red giants span a range in T$_{\rm eff}$ from 3400K to 4540K and log g from 0.5 to 2.1, covering
a significant part of the effective temperature - surface gravity range being observed in the APOGEE survey.

Chemical abundances are determined for 16 elements, along with the minor carbon isotope, $^{13}$C.
Comparisons of the abundances derived here with those from previously published studies of four of
the target stars here, mostly from visual-wavelength high-resolution spectra, find abundance differences
of $\sim$0.1 dex.  The APOGEE wavelength window with its derived linelist and model atmosphere grid,
coupled to spectrum synthesis modelling, can be used to analyze high-resolution H-band spectra of
red giants to probe both Galactic chemical evolution in stellar populations, as well as internal
red giant nucleosynthesis and mixing.

\acknowledgments
\section{Acknowledgments}

We thank Ken Hinkle, for helping to make the archival FTS spectral available and to Luan Ghezzi for
useful comments on an initial draft of this paper.  VS acknowledges
partial support for this research from the National Science Foundation (AST1109888) and CAPES, Brazil.  
This publication makes use of data products from the Two Micron All Sky Survey, which is a joint project of the
University of Massachusetts and the Infrared Processing and Analysis Center/California Institute of Technology, 
funded by the National Aeronautics and Space Administration and the National Science Foundation.  This research 
has made use of the SIMBAD database, operated at CDS, Strasbourg, France.

Funding for SDSS-III has been provided by the Alfred
P. Sloan Foundation, the Participating Institutions, the
National Science Foundation, and the U.S. Department
of Energy Office of Science. The SDSS-III web site is
http://www.sdss3.org/.

SDSS-III is managed by the Astrophysical Research
Consortium for the Participating Institutions of the
SDSS-III Collaboration including the University of Ari-
zona, the Brazilian Participation Group, BrookhavenNa-
tional Laboratory, University of Cambridge, Carnegie
Mellon University, University of Florida, the French
Participation Group, the German Participation Group,
Harvard University, the Instituto de Astrofisica de Ca-
narias, the Michigan State/Notre Dame/JINA Participa-
tion Group, Johns Hopkins University, Lawrence Berke-
ley National Laboratory, Max Planck Institute for Astro-
physics, New Mexico State University, New York Univer-
sity, Ohio State University, Pennsylvania State Univer-
sity, University of Portsmouth, Princeton University, the
Spanish Participation Group, University of Tokyo, Uni-
versity of Utah, Vanderbilt University, University of Vir-
ginia, University of Washington, and Yale University.



{\it Facilities:} \facility{NOAO}, \facility{Mayall 4m (FTS)}.

\clearpage



\begin{figure}
\epsscale{.80}
\plotone{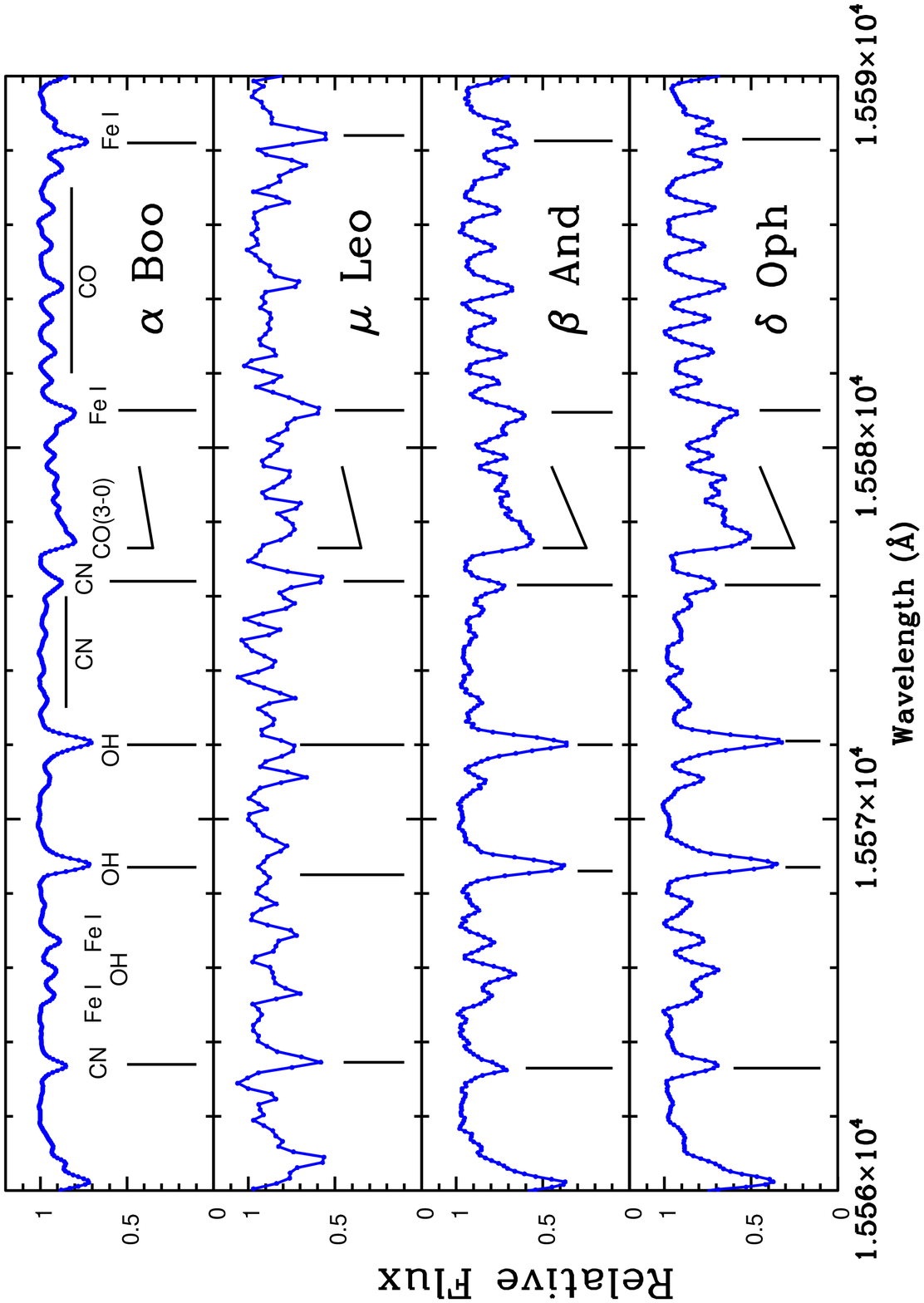}
\caption{ Sample spectra of red giants in the region near the $^{12}$C$^{16}$O (3--0) bandhead. 
These stars span a range in temperature, from T$_{eff}$ = 4550 K ($\mu$ Leo)
down to T$_{\rm eff}$ = 3825 K ($\beta$ And) and in metallicity, with $\alpha$ Boo
having [Fe/H] =-0.52, while $\mu$ Leo has [Fe/H]=+0.31 dex. These panels illustrate
both the quality of the spectra and the variation of molecular and atomic line
absorption with primarily T$_{\rm eff}$.
\label{fig1}}
\end{figure}

\clearpage


\begin{figure}
\plotone{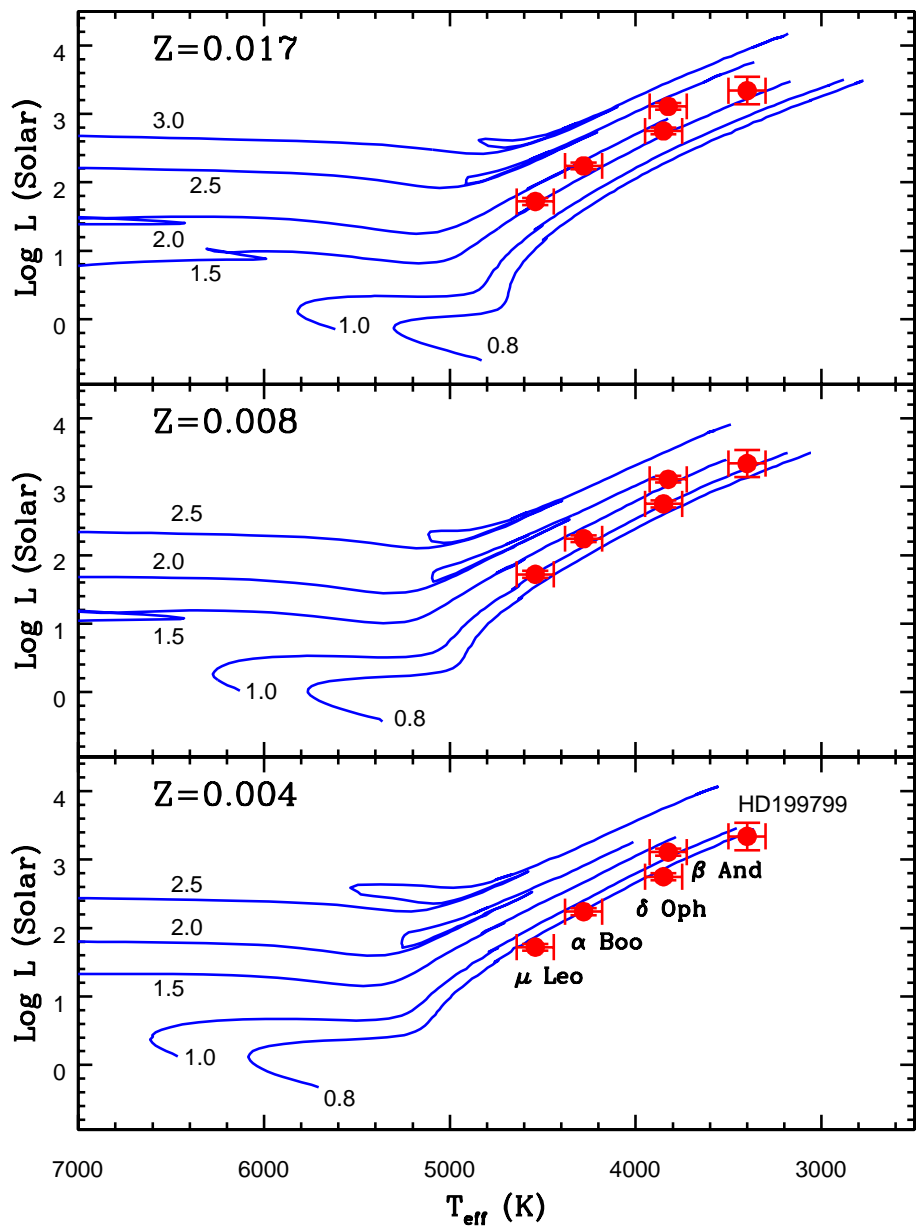}
\caption{The positions of the field red giants in the Log (L/L$_{\odot}$) - T$_{\rm eff}$ plane,
which is a modified HR-diagram. The continous curves represent stellar evolution
tracks for different mass stars with each panel representing a different heavy-element
mass fraction (Z); this is equivalent to overall stellar metallicity with the Sun
having Z=0.0145 (Lodders 2010). The model tracks are used to set log g for each red giant
studied here. 
\label{fig2}}
\end{figure}

\clearpage

\begin{figure}
\plotone{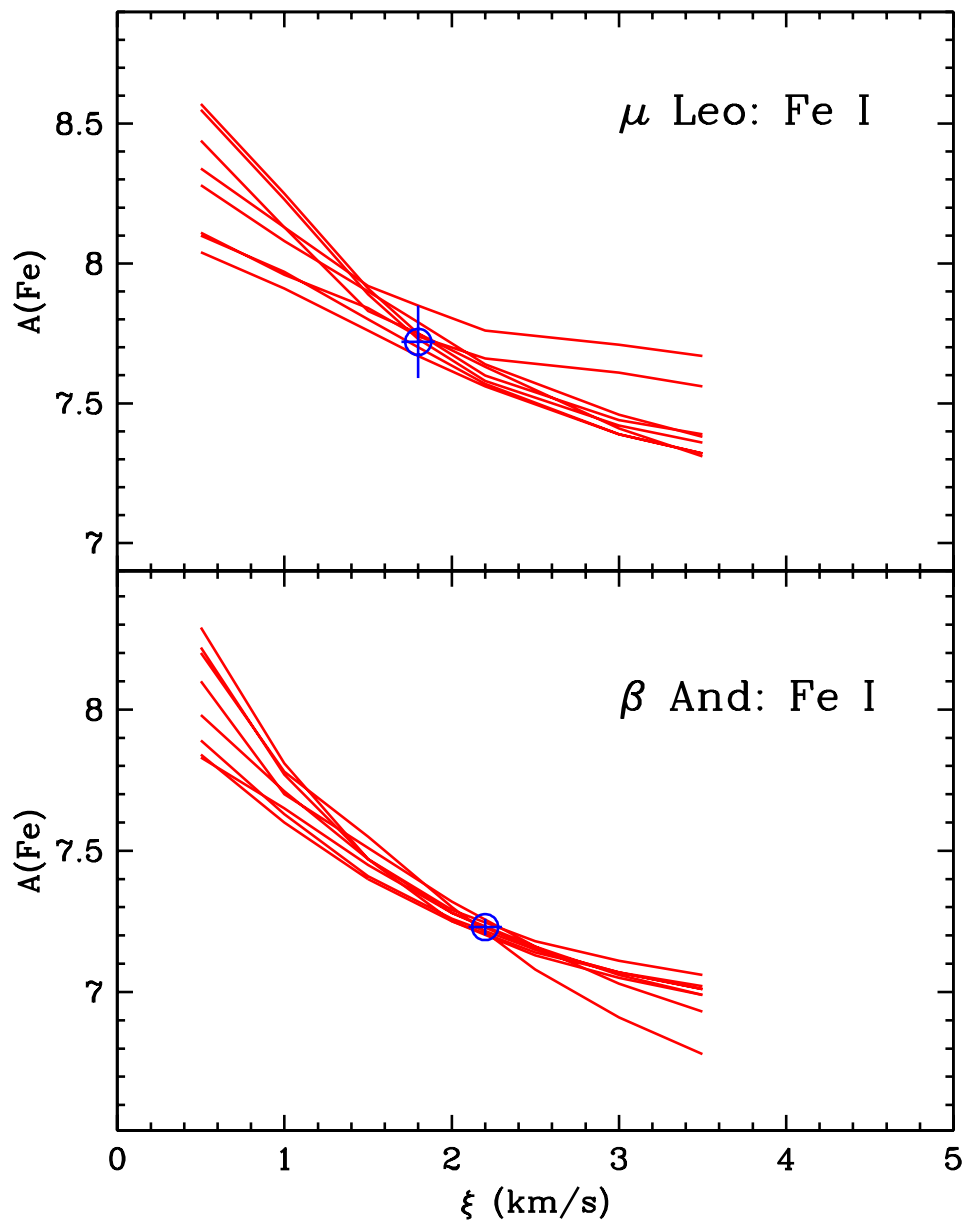}
\caption{The individual Fe I abundances as a function of microturbulent velocity
for $\mu$ Leo (top panel) and $\beta$ And (bottom panel). Each curve represents 
a single Fe I line.
Stronger lines are more sensitive to changes in $\xi$ than weaker lines. The value of $\xi$ that yields
no significant abundance differences as a function of reduced line-strength 
(log W$_{\lambda}$/$\lambda$) is taken as the characteristic microturbulent velocity
for the star. The derived microturbulent velocity as displayed here also
corresponds to the smallest scatter in A(Fe).
\label{fig3}}
\end{figure}

\clearpage

\begin{figure}
\epsscale{0.70}
\plotone{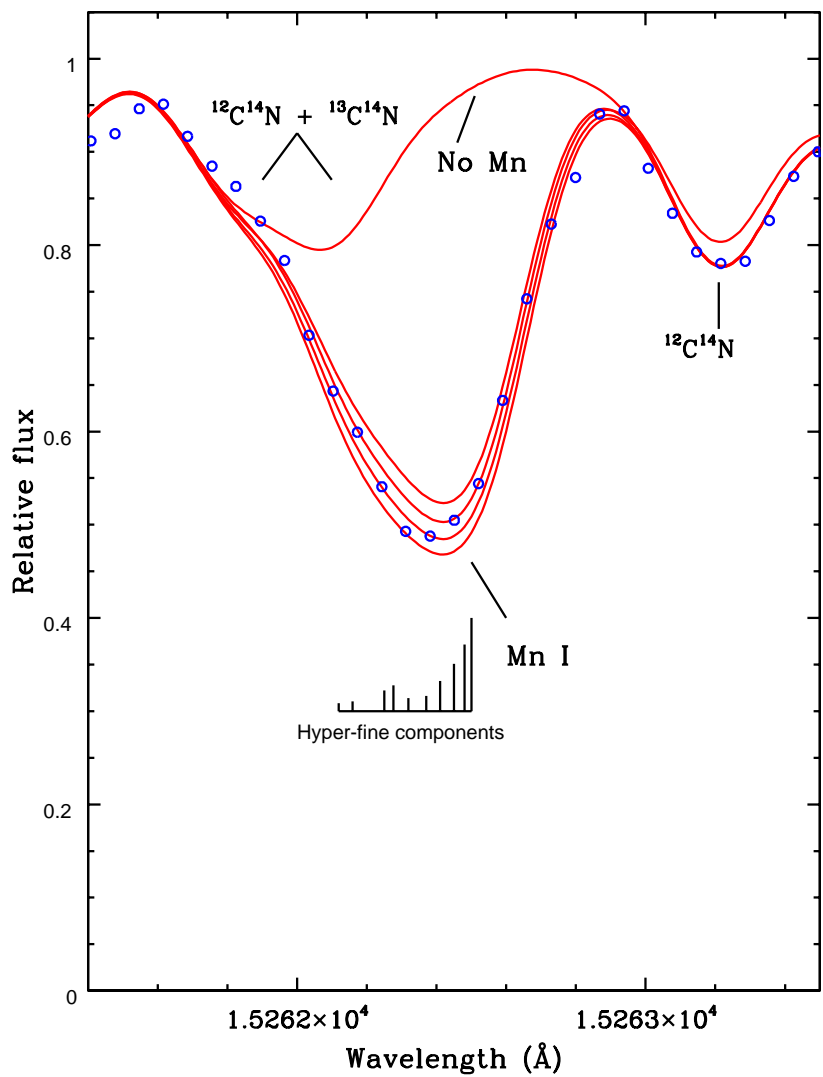}
\caption{A comparison of the observed spectrum near the Mn I line at 15262\AA 
in $\delta$ Oph (blue open circles) with synthetic spectra which were 
computed for Mn abundances separated in intervals of 0.1 dex. 
This Mn I transition consists of a number of hfs components of differing 
wavelengths, as shown, with the vertical length of
each component proportional to the gf-value of that component transition.
Note that Mn I is blended with a CN feature containing both $^{12}$C$^{14}$N and $^{13}$C$^{14}$N lines,
thus the value of  $^{12}$C/$^{13}$C ratio must be included. Both the hfs structure and
CN blending illutrates the need for the APOGEE analysis to be based on spectrum
synthesis with an accurate linelist. 
\label{fig4}}
\end{figure}

\clearpage

\begin{figure}
\epsscale{0.80}
\plotone{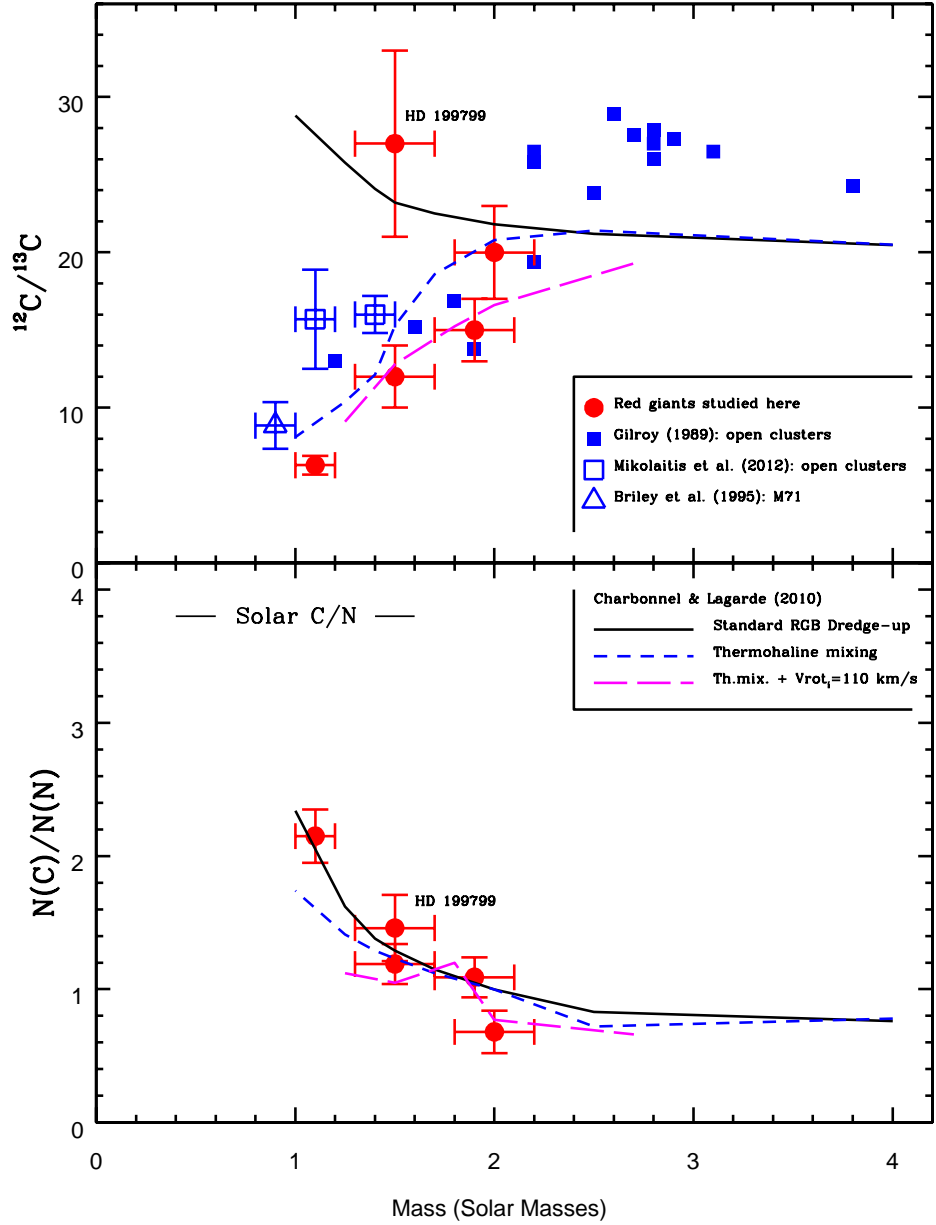}
\caption{The behavior of the nuclei $^{12}$C, $^{13}$C, and $^{14}$N with stellar mass on the
RGB as characterized by $^{12}$C/$^{13}$C and C/N (where C=$^{12}$C+$^{13}$C and N=$^{14}$N).
The relative abundances of these nuclei are altered by H-burning via the CN-cycle
and convective mixing on the RGB dredges up the processed abundances.  The TP-AGB star HD 199799
is identified, as it has increased significantly its $^{12}$C abundance from He-burning and third dredge-up
and has thus increased its $^{12}$C/$^{13}$C relative to when it was on the RGB.
\label{fig5}}
\end{figure}

\clearpage

\begin{figure}
\epsscale{0.75}
\plotone{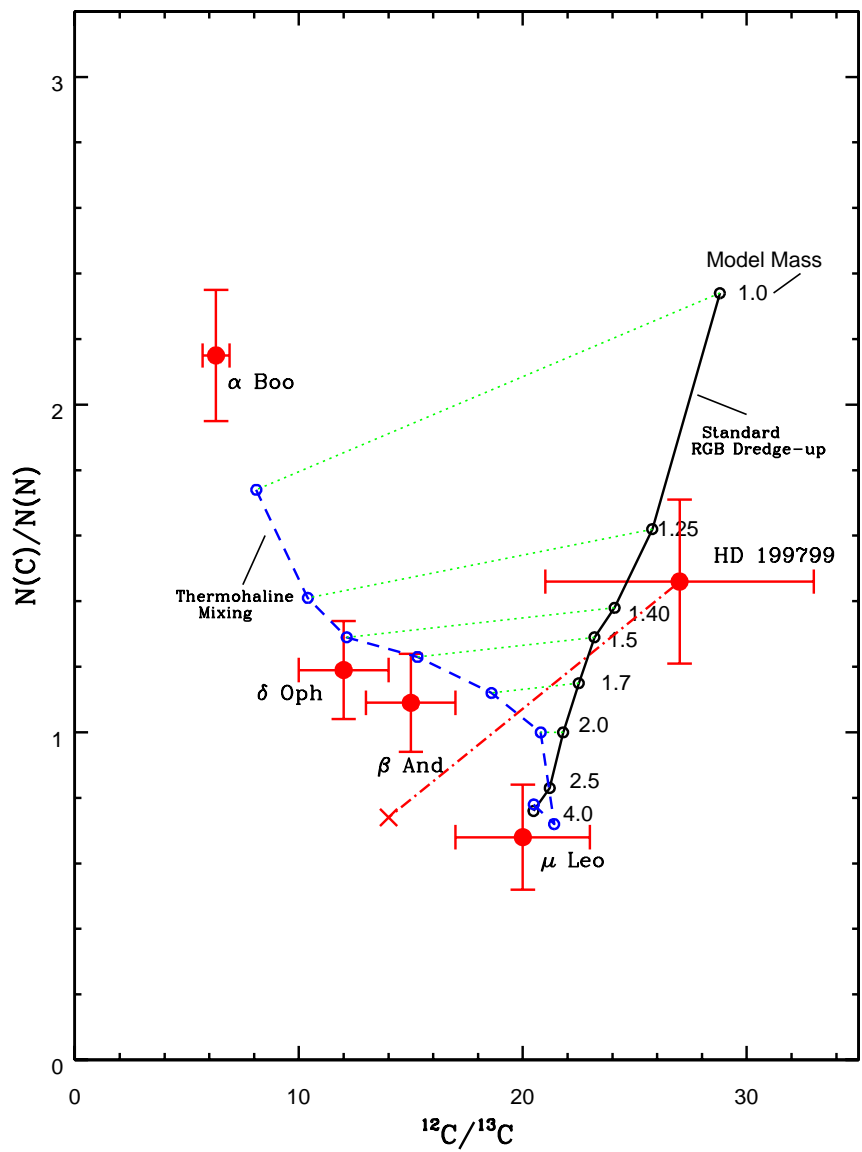}
\caption{Observed and model predicted behavior of N(C)/N(N) versus $^{12}$C/$^{13}$C;
the model curves show so-called standard dredge-up (solid line curve) and 
a thermohaline mixing model as a function of stellar mass from 
Charbonel \& Lagarde (2010; dashed line curve). 
Standard dredge-up does not predict the observed behavior of N(C)/N(N) as a
function of $^{12}$C/$^{13}$C and some type of additional mixing process is needed.  The
TP-AGB star HD 199799 is identified here as it has dredged-up enough $^{12}$C to have
approximately doubled its abundance relative to its abundance on the RGB; the dashed-dotted
line moves HD 199799 to its approximate position when it was evolving along the RGB.
\label{fig6}}
\end{figure}

\clearpage

\begin{figure}
\plotone{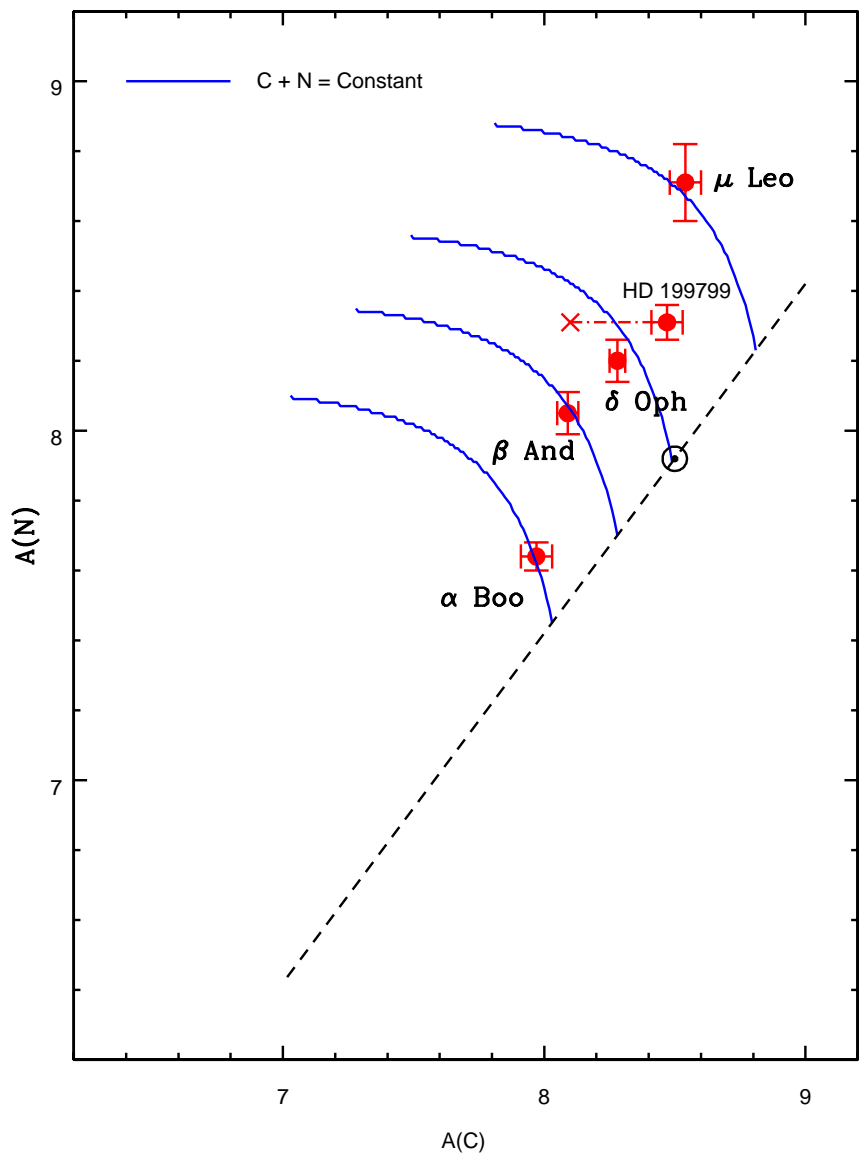}
\caption{Testing if total N(C) + N(N) is conserved, as would be expected for pure CN-cycle
dredge-up. The dashed line represents metallicity scaled values of A(C) and A(N)
with a solar C/N ratio. The blue curves map constant values of C+N scaled to
the metallicity (taken to be [Fe/H]) of each star. Clearly, initial (main-sequence) values
of [C+N/Fe]=0, followed by simple CN-cycle dredge-up, are adequate representations
of what is observed in these field RGB stars.  The TP-AGB star HD 199799 has increased it
$^{12}$C abundance by approximately 0.35 dex and this shift in A(C) is shown by the
dashed-dotted line.
\label{fig7}}
\end{figure}

\clearpage

\begin{figure}
\plotone{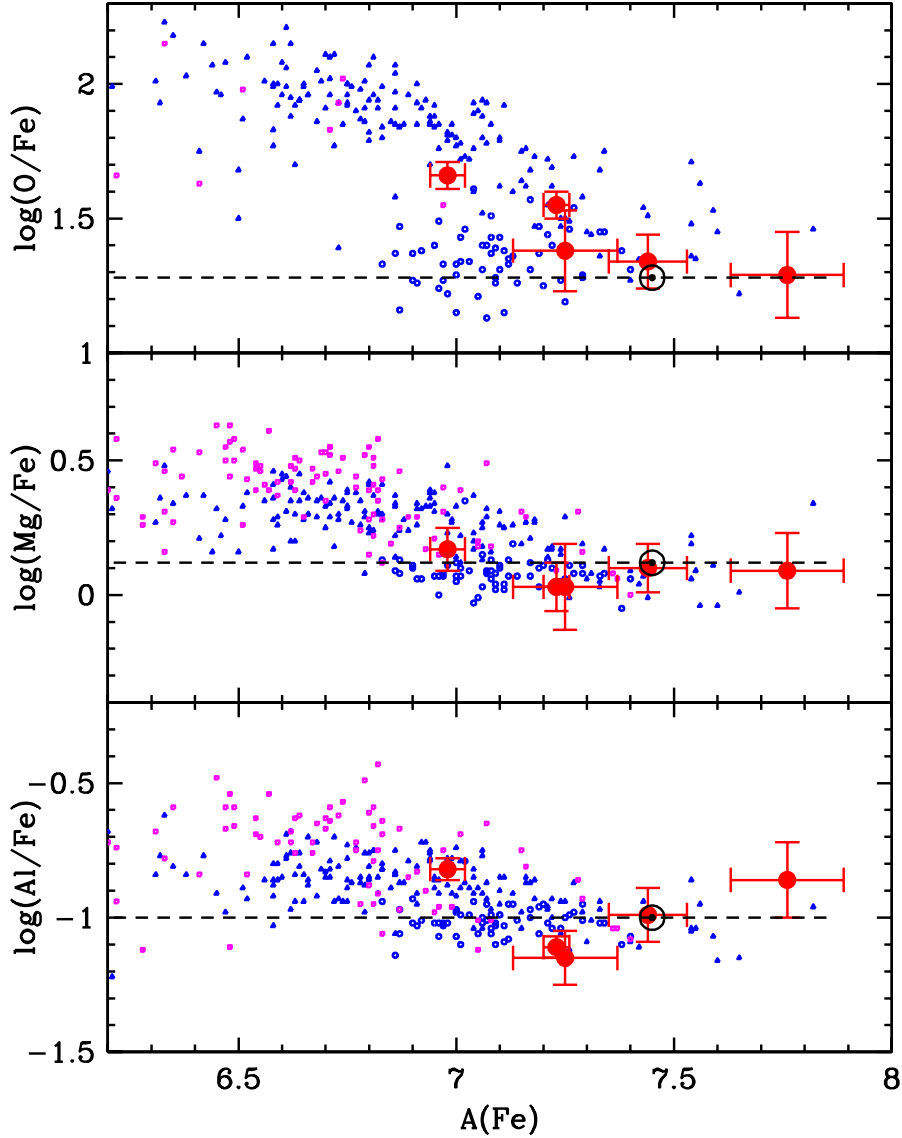}
\caption{The abundances of oxygen, magnesium, and aluminium, relative to iron, shown as
log(N(X)/N(H)) versus A(Fe). The 5 stars studied here are represented by the
filled red circles. 
The small blue symbols are results from Reddy et al. (2003; 2006) and the lower metallicity
small majenta symbols are from Fulbright (2002) and Johnson (2003). 
The solar position is indicated with the dashed line showing the solar abundance ratio.
The abundances derived here using the APOGEE linelist are well mapped into the general 
trends observed for Galatic field populations.  The error bars are taken from the standard deviations
in Table 6.  Errors for the ratios are quadrature sums of the element in-question and Fe I
values of $\sigma$.
\label{fig8}}
\end{figure}

\clearpage

\begin{figure}
\plotone{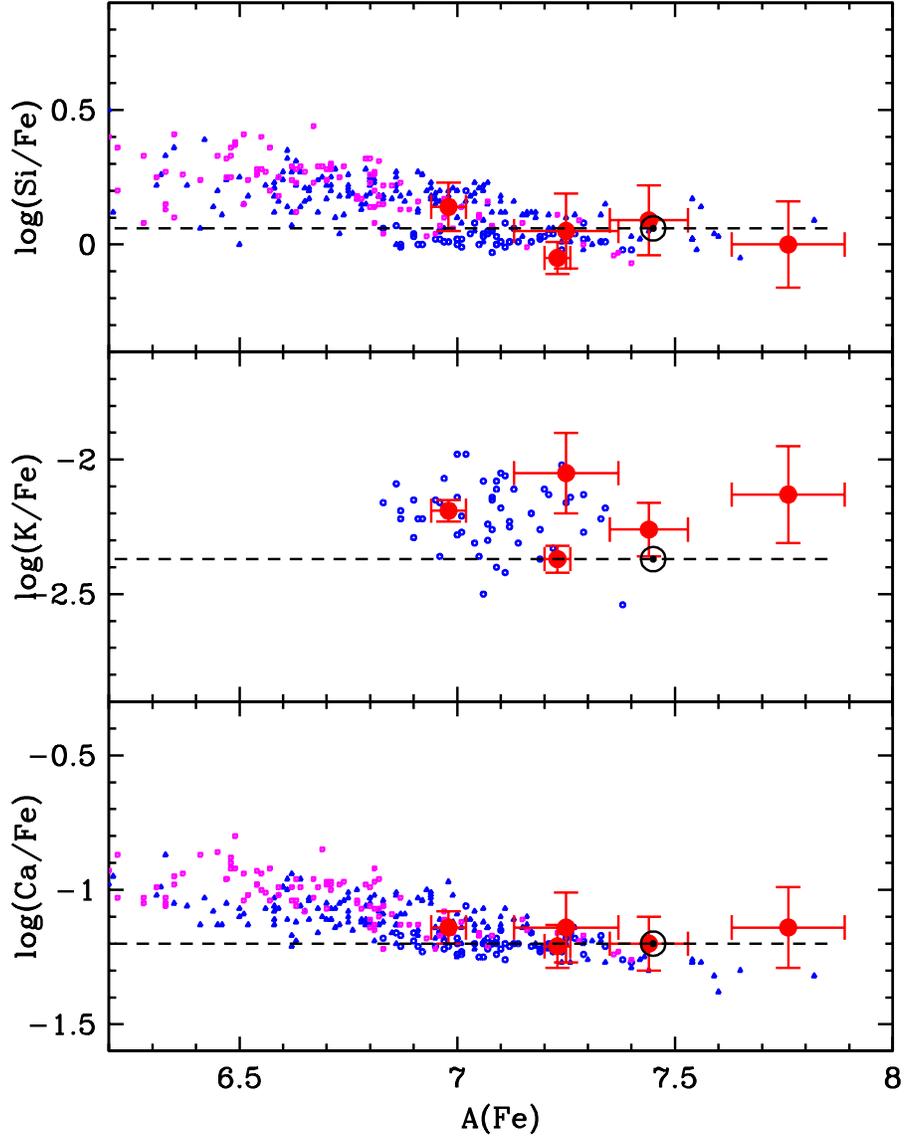}
\caption{The abundances of silicon, potassium, and calcium, relative to iron, versus A(Fe).  The
plotting format and symbols are the same as in Figure 8.
\label{fig9}}
\end{figure}

\clearpage

\begin{figure}
\plotone{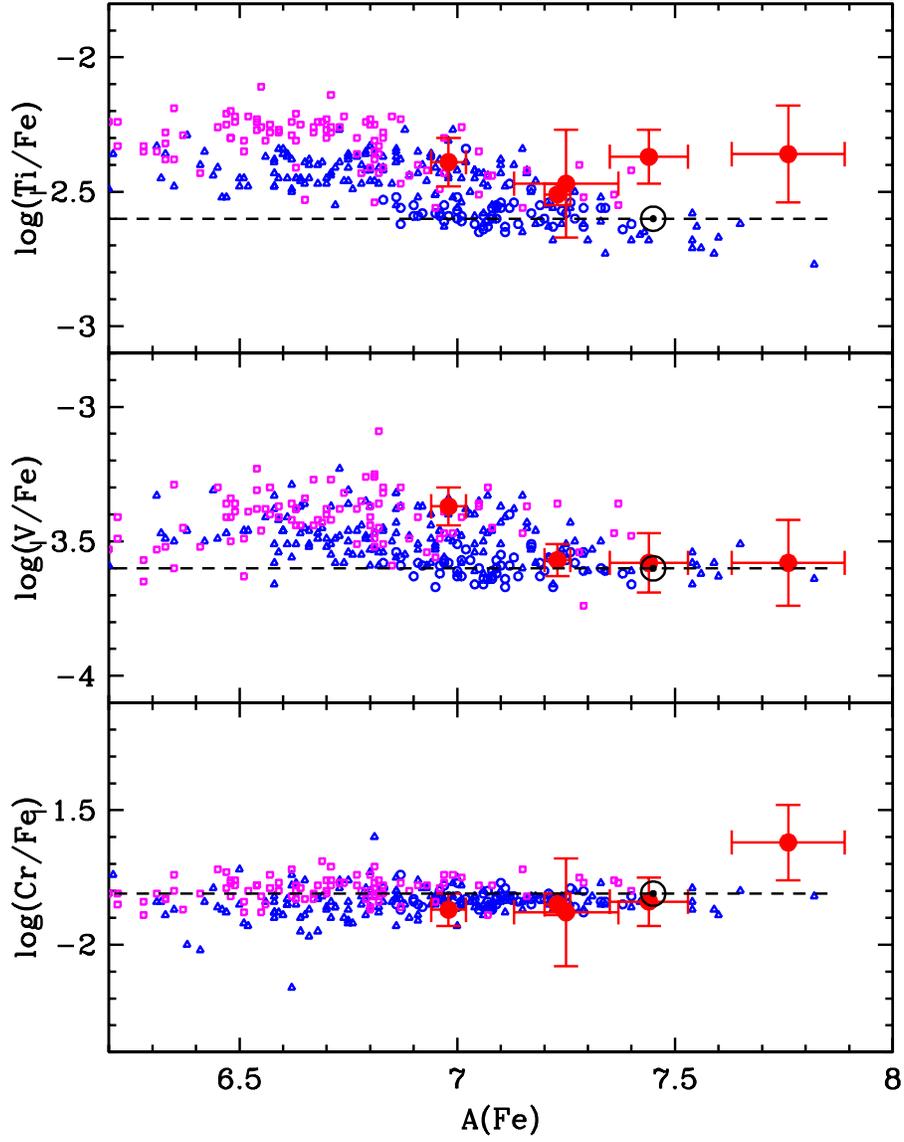}
\caption{The abundances of titanium, vanadium, and chromium, relative to iron, versus A(Fe).
The plotting format and symbols are the same as in Figure 8.
\label{fig10}}
\end{figure}

\clearpage

\begin{figure}
\plotone{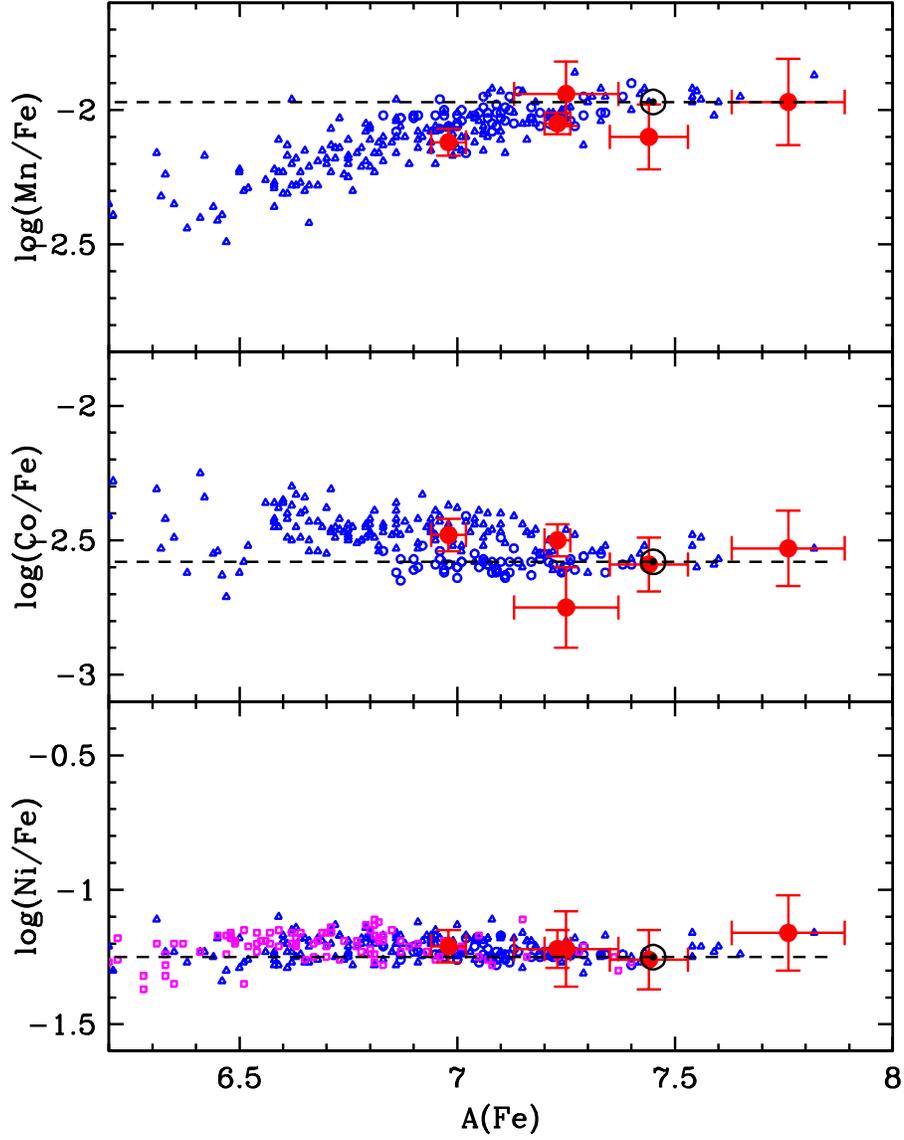}
\caption{The abundances of manganese, cobalt, and nickel, relative to iron, versus A(Fe).
The plotting format and symbols are the same as in Figure 8.
\label{fig11}}
\end{figure}

\clearpage

\begin{figure}
\plotone{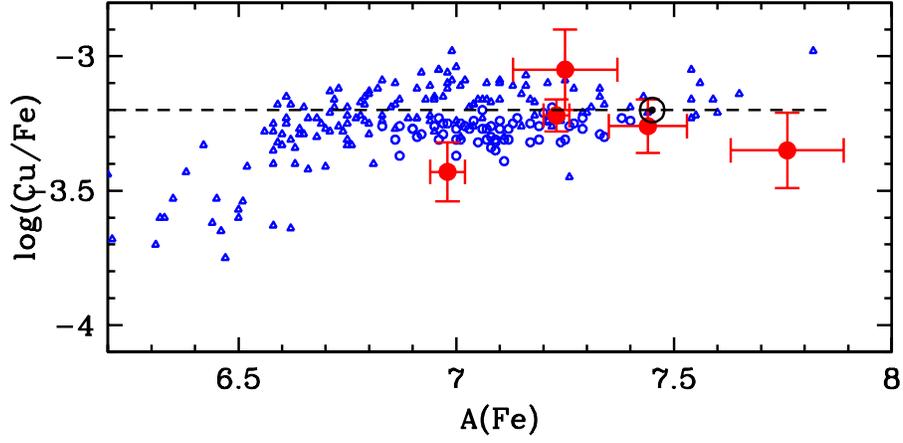}
\caption{The abundance of copper relative to iron versus A(Fe).  The plotting format and symbols
are the same as in Figure 8.
\label{fig12}}
\end{figure}

\clearpage









\clearpage

\begin{deluxetable}{ccccccc}
\tabletypesize{\scriptsize}
\tablecaption{Red Giant Standard Stars: Observed Properties}
\tablewidth{0pt}
\tablehead{
\colhead{Star} & \colhead{HR} & \colhead{SpT} & \colhead{$\pi$ (mas)\tablenotemark{a}} & \colhead{d(pc)} &
\colhead{J-K$_{\rm S}$\tablenotemark{b}} &
\colhead{M$_{\rm bol}$\tablenotemark{c}} 
}
\startdata
$\beta$ And  & 337   & M0III & 16.4$\pm$0.8 & 61$\pm$3          & 0.98 & -3.06$\pm$0.11 \\
$\mu$ Leo    & 3905 & K2III &  26.3$\pm$0.2 & 38$\pm$1           & 0.68 &   0.41$\pm$0.05 \\
$\alpha$ Boo & 5340 & K2III & 88.8$\pm$0.5 &  11.3$\pm$0.1     & 0.79 & -0.92$\pm$0.03 \\
$\delta$ Oph & 6056 & M0III & 19.1$\pm$0.2 & 52.5$\pm$0.5     & 0.97 & -2.16$\pm$0.05 \\
HD 199799    &  ...     & M2S  &  2.6$\pm$0.6  & 386$\pm$92       & 1.25 & -3.64$\pm$0.54 \\
\enddata
\tablenotetext{a}{Parallax from van Leeuwen (2007)}
\tablenotetext{b}{Johnson (1965) transformed to the 2MASS system using Carpenter (2001)}
\tablenotetext{c}{Calculated from M$_{\rm K}$ with bolometric corrections from Bessell et al. (1998)}
\end{deluxetable}


\clearpage

\begin{deluxetable}{ccccccc}
\tabletypesize{\scriptsize}
\tablecaption{Red Giant Standard Stars: Derived Parameters}
\tablewidth{0pt}
\tablehead{
\colhead{Star} & \colhead{Log(L/L$_{\odot}$)} & \colhead{M/M$_{\odot}$} & \colhead{T$_{\rm eff}$ (K)} &
\colhead{Log g (cm-s$^{-2}$)} & \colhead{$\xi$ (km-s$^{-1}$)} & \colhead{[Fe/H]\tablenotemark{a}}
}
\startdata
$\beta$ And      &  3.11$\pm$0.05  &  1.9$\pm$0.2  &  3825$\pm$75  &  0.9$\pm$0.1  &  2.20$\pm$0.10  &  -0.2  \\
$\mu$ Leo        &  2.68$\pm$0.03  &  2.0$\pm$0.3  &  4550$\pm$50  &  2.1$\pm$0.1  &  1.80$\pm$0.15  &  +0.3  \\
$\alpha$ Boo    &  2.24$\pm$0.02  &  1.1$\pm$0.2  &  4275$\pm$50  &  1.7$\pm$0.1  &  1.85$\pm$0.05  &  -0.4  \\
$\delta$ Oph    &  2.75$\pm$0.03  &  1.5$\pm$0.2  &  3850$\pm$75  &  1.2$\pm$0.1  &  1.90$\pm$0.10  &  +0.0  \\
 HD 199799      &  3.34$\pm$0.22  &  1.5$\pm$0.2  &  3400$\pm$75  &  0.5$\pm$0.2  &  2.40$\pm$0.20  &  -0.2  \\
\enddata

\tablenotetext{a}{[Fe/H] represents the overall metallicity used in the final model atmosphere.  Final Fe abundances
are given in Table 6}

\end{deluxetable}

\clearpage

\begin{deluxetable}{cccccccc}
\tabletypesize{\scriptsize}
\tablecaption{Fe I Lines used in the Abundance Determinations}
\tablewidth{0pt}
\tablehead{
\colhead{$\lambda$ (\AA)} & \colhead{$\chi$ (eV)} & \colhead{log gf} & \colhead{$\alpha$ Boo} &
\colhead{$\beta$ And} & \colhead{$\delta$ Oph} & \colhead{$\mu$ Leo} & \colhead{HD 199799}
}
\startdata
 15194.492  & 2.223  & -4.779  & 7.04  & 7.22  & 7.37  & 7.83  & 7.05  \\
 15207.526 & 5.385  & +0.080  & 7.04  & 7.29  & 7.40  & bl$^{a}$  & 7.20  \\
 15395.718 & 5.620  & -0.341  & 6.96  & 7.27  &    bl$^{a}$   & 7.88  & 7.42  \\
 15490.339 & 2.198  & -4.807  & 6.99  & 7.23  & 7.55  & 7.76  & 7.50  \\
 15648.510 & 5.426  & -0.701  & 6.97  & 7.20  & 7.35  & 7.75  & 7.03  \\
 15964.867 & 5.921  & -0.128  & 6.93  & 7.22  & 7.35  & 7.65  & 7.33  \\
 16040.657 & 5.874  & +0.066  & 6.94  & 7.20  &   bl$^{a}$   & 7.69  & 7.14  \\
 16153.247 & 5.351  & -0.743  & 6.94  & 7.22  & 7.53  & 7.75  & 7.32  \\
 16165.032 & 6.319  & +0.723  & 6.98  & 7.24  & 7.56  & 7.73  & 7.30  \\   
\enddata

\tablenotetext{a}{bl: Feature too blended to use in cooler or very metal-rich red giants.}
\end{deluxetable}

\clearpage

\begin{deluxetable}{ccccccc}
\tabletypesize{\scriptsize}
\tablecaption{Molecular Lines and Features used for $^{12}$C, $^{13}$C, $^{14}$N, and $^{16}$O}
\tablewidth{0pt}
\tablehead{
\colhead{Molecular Lines} & \colhead{$\lambda$--interval (\AA)} & \colhead{$\alpha$ Boo} & \colhead{$\beta$ And} &
\colhead{$\delta$ Oph} & \colhead{$\mu$ Leo} & \colhead{HD 199799}
}
\startdata
 {\bf $^{12}$C from $^{12}$C$^{16}$O lines} & & & & & & \\
 (3-0) V-R       & 15578 -- 15586 & 7.86 & 8.05 & 8.27 &  w$^{a}$ & 8.51 \\
 (4-1) V-R       & 15774 -- 15787 & 7.97 & 8.11 & 8.24 & 8.46 & 8.51 \\
 (5-2) V-R       & 15976 -- 16000 & 7.90 & 8.07 & 8.23 & 8.56 & 8.38 \\
 (6-3) V-R       & 16183 -- 16196 & 7.90 & 8.03 & 8.22 & 8.54 & 8.43 \\
{\bf $^{12}$C/$^{13}$C ratios from $^{13}$C$^{16}$O and $^{13}$C$^{14}$N lines} & & & & & & \\
$^{13}$C$^{16}$O(3-0) V-R    & 15922 -- 15926 & 7 & 17 & 14 & w$^{a}$ & 30 \\
$^{13}$C$^{16}$O(4-1) V-R    & 16120 -- 16125 & 6 & 15 & 11 & w$^{a}$ & 32 \\
$^{13}$C$^{16}$O(6-3) V-R    & 16740 -- 16747 & 6 & 13 & 11 & w$^{a}$ & 20 \\
$^{13}$C$^{14}$N                   & 15314 -- 15315 & w$^{a}$ & w$^{a}$ & w$^{a}$ & 22 & w$^{a}$ \\
$^{13}$C$^{14}$N                   & 15354 -- 15356 & w$^{a}$ & w$^{a}$ & w$^{a}$ & 17 & w$^{a}$ \\
 {\bf $^{16}$O from $^{16}$OH lines} & & & & & & \\
 (2-0) P$_{1}$ 9.5    & 15277 -- 15282  & 8.55 & 8.83 & 8.78 & 9.10 & 8.80 \\
 (3-1) P$_{2}$ 3.5    & 15390 -- 15392  & 8.54 & 8.78 & 8.76 & 8.93 & 8.60 \\
 (3-1) P$_{2}$ 4.5    & 15504 -- 15507  & 8.48 & 8.71 & 8.73 & 9.15 & 8.58 \\
 (2-0) P$_{2}$ 11.5  & 15568 -- 15573  & 8.51 & 8.80 & 8.78 & 8.98 & 8.52 \\
 (3-1) P$_{2}$ 9.5    & 16189 -- 16193  & 8.47 & 8.78 & 8.81 & 9.07 & 8.63 \\
{\bf $^{14}$N from $^{12}$C$^{14}$N lines} & & & & & & \\
 (1-2) Q2 41.5         & 15260. & 7.61 & 8.10 & 8.23 & 8.83 & 8.31 \\
 (1-2) P2 34.5          & 15322. & 7.68 & 8.07 & 8.21 & 8.88 & 8.35 \\
 (1-2) R2 56.5         & 15397. & 7.65 & 8.06 & 8.26 & 8.63 & 8.28\\
 (0-1) R1 68.5         & 15332. & 7.63 & 7.95 & bl$^{b}$  & 8.68 & 8.35 \\
 (0-1) P2 49.5         & 15410. & 7.66 & 8.03 & 8.13 & 8.80 & bl$^{b}$  \\
 (0-1) Q2 59.5        & 15447. & 7.64 & 7.98 & 8.28 & 8.73 & bl$^{b}$  \\
 (0-1) Q1 60.5        & 15466. & 7.65 & 8.06 & 8.23 & 8.51 & bl$^{b}$  \\
 (1-2) P2 38.5         & 15472. & 7.71 & 8.13 & 8.11 & 8.68 & 8.25 \\
 (0-1) P1 51.5        & 15482. & 7.56 & 8.03 & 8.15 & 8.68  & 8.25 \\
\enddata

\tablenotetext{a}{w: Feature too weak to use.}
\tablenotetext{b}{bl: Feature too blended by nearby lines.}
\end{deluxetable}

\clearpage

\begin{deluxetable}{ccccccccc}
\tabletypesize{\scriptsize}
\tablecaption{Atomic Lines used and Derived Abundance}
\tablewidth{0pt}
\tablehead{
\colhead{Element} & \colhead{$\lambda$ ($\AA $)} & \colhead{$\chi$ (eV)} & \colhead{log gf} & \colhead{$\alpha$ Boo} &
\colhead{$\beta$ And} & \colhead{$\delta$ Oph} & \colhead{$\mu$ Leo} & \colhead{HD 199799}
}
\startdata
 {\bf Mg I} & 15740.716 & 5.931 & -0.262 & 7.15 & 7.25 & 7.53 & 7.80 & 7.29 \\
                 & 15748.9     & 5.932 & +0.276 & 7.09 &  s$^{a}$ & 7.57 & s$^{a}$  & 7.23 \\
                 & 15765.8     & 5.933 & +0.504 & 7.04 &  s$^{a}$ &  s$^{a}$ & 7.87 & bl$^{b}$  \\
                 & 15879.5     & 5.946 & -1.248 & 7.13 & 7.20 & 7.45 & 7.82 & bl$^{b}$  \\
                 & 15886.2     & 5.946 & -1.555 & 7.22 & 7.21 & 7.53 &  bl$^{b}$ & 7.18 \\
                 & 15889.485 & 5.946 & -2.013 &  bl$^{b}$ & 7.30 & 7.55 & 7.90 & w$^{c}$  \\
                 & 15954.477 & 6.588 & -0.807 & 7.25 & 7.36 & 7.61 &  bl$^{b}$ & 7.45 \\
 {\bf Al I}   & 16718.957 & 4.085 & +0.290 & 6.15 & 6.15 &  s$^{a}$ & 6.85 & bl$^{b}$  \\
                 & 16763.359 & 4.087 & -0.524 & 6.16 & 6.09 & 6.45 & 6.94 & 6.10 \\
 {\bf Si I}   & 15361.161 & 5.954 & -1.925 & 6.99 &  bl$^{b}$ &  bl$^{b}$ &  bl$^{b}$ & w$^{c}$  \\
                 & 15376.831 & 6.222 & -0.649 & 7.18 & 7.17 & 7.57 & 7.66 & w$^{c}$  \\
                 & 15833.602 & 6.222 & -0.168 & 6.93 & 6.95 & 7.28 & 7.72 & 7.20 \\
                 & 15960.063 & 5.984 & +0.107 & 7.07 & 7.14 & 7.43 & 7.88 & w$^{c}$  \\
                 & 16060.009 & 5.954 & -0.566 & 7.21 & 7.25 & 7.59 & 7.74 & 7.33 \\
                 & 16094.787 & 5.964 & -0.168 & 7.11 & 7.06 & 7.46 & 7.66 & 7.32 \\
                 & 16215.670 & 5.964 & -0.665 & 7.15 & 7.23 & 7.62 &  bl$^{b}$ & bl$^{b}$  \\
                 & 16680.770 & 5.984 & -0.140 & 7.11 & 7.25 & 7.62 & 7.83 & 7.35 \\
                 & 16828.159 & 5.984 & -1.102 & 7.13 & 7.17 & 7.45 & 7.85 & bl$^{b}$  \\
 {\bf K I}    & 15163.067 & 2.670 & +0.524 & 4.81 & 4.82 & 5.18 & 5.63 & 5.20 \\
                 & 15168.376 & 2.670 & +0.347 & 4.77 & 4.89 &  bl$^{b}$ &  bl$^{b}$ & bl$^{b}$  \\
 {\bf Ca I}  & 16136.823 & 4.531 & -0.552 & 5.75 & 6.09 & 6.23 & 6.69 & 6.06 \\
                 & 16150.763 & 4.532 & -0.229 & 5.88 & 5.95 & 6.21 & 6.53 & 6.07 \\
                 & 16155.236 & 4.532 & -0.619 & 5.82 & 6.02 & 6.30 & 6.61 & 6.16 \\
                 & 16157.364 & 4.554 & -0.208 & 5.90 & 6.02 & 6.23 & 6.63 & 6.16 \\
 {\bf Ti I}   & 15543.756 & 1.879 & -1.120 & 4.53 & 4.74 & 5.05 & 5.58 & 4.74 \\
                 & 15602.842 & 2.267 & -1.643 & 4.66 & 4.74 & 5.03 & 5.43 & bl$^{b}$  \\
                 & 15698.979 & 1.887 & -2.060 & 4.52 & 4.69 & 5.03 & 5.13 & bl$^{b}$  \\
                 & 15715.573 & 1.873 & -1.250 & 4.57 & 4.71 & 5.14 &  bl$^{b}$ & 4.70 \\
                 & 16635.161 & 2.345 & -1.807 & 4.65 &  bl$^{b}$ & 5.10 & 5.46 & 4.90 \\
 {\bf V I}    & 15924.       & 2.138 & -1.175 & 3.61 & 3.66 & 3.86 & 4.18 & bl$^{b}$  \\ 
 {\bf Cr I}  & 15680.063 & 4.697 & +0.179 & 5.08 & 5.38 & 5.62 & 6.11 & 5.20 \\
                & 15860.214 & 4.697 & -0.012 & 5.14 & 5.39 & 5.57 & 6.16  & 5.54  \\
 {\bf Mn I} & 15159.       & 4.889 & +0.544 & 4.88 & 5.19 &  bl$^{b}$ & 5.82 & bl$^{b}$  \\
                 & 15217.       & 4.889 & +0.414 & 4.91 & 5.26 & 5.47 & 5.83 & 5.32 \\
                 & 15262.       & 4.889 & +0.345 & 4.78 & 5.08 & 5.21 & 5.71 & 5.30 \\
 {\bf Co I}  & 16757.7     & 3.409 & -1.230 & 4.50 & 4.73 & 4.85 & 5.23 & 4.50 \\
 {\bf Ni I}   & 15605.680 & 5.299 & -0.376 & 5.76 & 6.08 & 6.18 & 6.69 & 6.05 \\
                 & 15632.654 & 5.305 & -0.106 & 5.82 & 5.99 & 6.20 & 6.65 & 6.02 \\
                 & 16584.439 & 5.299 & -0.528 & 5.71 & 6.00 & 6.21 & 6.58 & 6.12 \\
                 & 16589.295 & 5.469 & -0.600 & 5.81 & 6.03 & 6.23 & 6.72 & 5.98 \\
                 & 16673.711 & 6.029 & +0.317 & 5.72 &  bl$^{b}$ &  bl$^{b}$ & 6.39 & 6.12 \\
                 & 16815.471 & 5.305 & -0.606 & 5.79 & 5.94 & 6.06 & 6.58 & 5.95 \\
                 & 16818.760 & 6.039 & +0.311 & 5.77 &  bl$^{b}$ & 6.22 &  bl$^{b}$ & 5.95 \\
 {\bf Cu I}  & 16005.7     & 5.348 & -0.131 & 3.55 & 4.01 & 4.18 & 4.41 & 4.20 \\
\enddata

\tablenotetext{a}{s: Feature too strong to use.}
\tablenotetext{b}{bl: Feature too blended by nearby lines.}
\tablenotetext{c}{w: Feature too weak.}
\end{deluxetable}

\clearpage

\begin{deluxetable}{cccccc}
\tabletypesize{\scriptsize}
\tablecaption{Chemical Abundances}
\tablewidth{0pt}
\tablehead{
\colhead{Element} & \colhead{$\alpha$ Boo} & \colhead{$\beta$ And} & \colhead{$\delta$ Oph} &
\colhead{$\mu$ Leo} & \colhead{HD 199799}
}
\startdata
   Fe           & 6.98$\pm$0.04 & 7.23$\pm$0.03 & 7.44$\pm$0.09 & 7.76$\pm$0.08 & 7.25$\pm$0.12 \\
 $^{12}$C  & 7.96$\pm$0.05 & 8.06$\pm$0.03 & 8.24$\pm$0.02 & 8.52$\pm$0.05 & 8.46$\pm$0.06 \\
 $^{12}$C/$^{13}$C  & 6.3$\pm$0.6  & 15$\pm$2   & 12$\pm$2    & 20$\pm$3         & 27$\pm$6      \\
 $^{14}$N  & 7.64$\pm$0.04 & 8.05$\pm$0.06 & 8.20$\pm$0.06 & 8.71$\pm$0.11 & 8.31$\pm$0.05 \\
 $^{16}$O  & 8.64$\pm$0.04 & 8.78$\pm$0.04 & 8.77$\pm$0.03 & 9.05$\pm$0.09 & 8.63$\pm$0.09 \\
   Mg          & 7.15$\pm$0.08 & 7.26$\pm$0.07 & 7.54$\pm$0.05 & 7.85$\pm$0.05 & 7.28$\pm$0.11 \\
   Al           & 6.16$\pm$0.01 & 6.12$\pm$0.03 & 6.45                   & 6.90$\pm$0.05 & 6.10                  \\
   Si           & 7.12$\pm$0.07 & 7.18$\pm$0.07 & 7.53$\pm$0.08  & 7.76$\pm$0.09 & 7.30$\pm$0.07 \\
    K           & 4.79$\pm$0.02 & 4.86$\pm$0.04 & 5.18                   & 5.63                   & 5.20                 \\
   Ca          & 5.84$\pm$0.07 & 6.02$\pm$0.06 & 6.24$\pm$0.04 & 6.62$\pm$0.07 & 6.11$\pm$0.05 \\
   Ti           & 4.59$\pm$0.07 & 4.72$\pm$0.02 & 5.07$\pm$0.05 & 5.40$\pm$0.19 & 4.78$\pm$0.10 \\
    V           & 3.61                  & 3.66                   & 3.86                   & 4.18                   &  ...                  \\
   Cr           & 5.11$\pm$0.03 & 5.38$\pm$0.01 & 5.60$\pm$0.03 & 6.14$\pm$0.03 & 5.37$\pm$0.17 \\
   Mn         & 4.86$\pm$0.07 & 5.18$\pm$0.09 & 5.34$\pm$0.13 & 5.79$\pm$0.07 & 5.31$\pm$0.01 \\
   Co         & 4.50                   & 4.73                   & 4.85                   & 5.23                   & 4.50                  \\
   Ni          & 5.77$\pm$0.04 & 6.01$\pm$0.05 & 6.18$\pm$0.06 & 6.60$\pm$0.12  & 6.03$\pm$0.07 \\
   Cu         & 3.55                   & 4.01                   & 4.18                  & 4.41                    & 4.20                  \\
\enddata

\end{deluxetable}

\clearpage 

\begin{deluxetable}{ccccccc}
\tabletypesize{\scriptsize}
\tablecaption{Abundance Sensitivity: T$_{\rm eff}$=4000K, Log g=1.3, $\xi$=2.0 km-s$^{-1}$, [m/H]=0.0}
\tablewidth{0pt}
\tablehead{
\colhead{Species} & \colhead{$\partial{A}/\partial{T}$ (+50K)} & \colhead{$\partial{A}/\partial{G}$ (+0.2 dex)} & 
\colhead{$\partial{A}/\partial{\xi}$ (+0.2 km-s$^{-1}$)} & \colhead{$\partial{A}/\partial{m}$ (+0.1 dex)} & 
\colhead{dA'} & \colhead{dA} 
}
\startdata
 Fe I   & +0.015 & +0.003 & -0.056 & +0.032 & 0.066 & 0.087 \\
 CO    & +0.032 & +0.072 & -0.011 & +0.041 & 0.089 & 0.167 \\
 CN    & -0.018  & +0.064 & -0.006 & +0.038 & 0.077 & 0.127 \\ 
 OH    & +0.124 & -0.022 & -0.079 & +0.111 & 0.186 & 0.363 \\
 Mg I  & +0.017 & -0.031 & -0.042 & +0.029 & 0.062 & 0.098 \\
 Al I   & +0.055 & -0.077 & -0.052 & +0.027 & 0.111 & 0.207 \\
 Si I   & +0.003 & -0.013 & -0.039 & +0.038 & 0.056 & 0.065 \\  
  K I   & +0.042 & -0.046 & -0.048 & +0.014 & 0.081 & 0.143 \\
 Ca I  & +0.045 & -0.046 & -0.032 & +0.017 & 0.074 & 0.143 \\
 Ti I   & +0.082 & -0.004 & -0.111 & +0.015 & 0.139 & 0.168 \\
  V I   & +0.076 & +0.009 & -0.054 & +0.011 & 0.094 & 0.132 \\
 Cr I   & +0.038 & -0.018 & -0.031 & +0.016 & 0.055 & 0.100 \\
 Mn I  & +0.032 & -0.057 & -0.056 & +0.041 & 0.095 & 0.163 \\
 Co I  & +0.026 & +0.033 & -0.075 & +0.034 & 0.092 & 0.136 \\
 Ni I   & -0.005 & +0.017 & -0.043 & +0.031 & 0.056 & 0.068 \\
 Cu I  & +0.001 & +0.032 & -0.013 & +0.019 & 0.039 & 0.044 \\
\enddata

\end{deluxetable}

\clearpage






\end{document}